\documentclass[preprint2,tighten,times]{aastex6}
\pdfoutput=1 
\usepackage{hyperref,verbatim}
\usepackage{amsmath,amstext,mathrsfs}
\usepackage[all]{hypcap} 
\usepackage{multirow}
\usepackage{color}
\usepackage{marginnote}

\begin{document}
\title{Tracing interstellar magnetic field using the velocity gradient technique \\in shock and self-gravitating media} 
\author{Ka Ho Yuen\altaffilmark{1,2}, A. Lazarian\altaffilmark{1}}
\email{kyuen2@wisc.edu, lazarian@astro.wisc.edu}
\altaffiltext{1}{Department of Astronomy, University of Wisconsin, Madison}
\altaffiltext{2}{Department of Physics, The Chinese University of Hong Kong}
\begin{abstract}
This study proceeds with the development of the technique employing velocity gradients that were identified in (\cite{GL17}, henceforth GL17) as a means of probing magnetic field in interstellar media. We demonstrate a number of practical ways on improving the accuracy of tracing magnetic fields in diffuse interstellar media using velocity centroid gradients (VCGs). Addressing the magnetic field tracing in super-Alfvenic turbulence we introduce the procedure of filtering low spatial frequencies, that enables magnetic field tracing in the situations when the kinetic energy of turbulent plasmas dominate its magnetic energy. We propose the synergic way of of using VCGs together with intensity gradients (IGs), synchrotron intensity gradients (SIGs) as well as dust polarimetry. We show that while the IGs trace magnetic field worse than the VCGs, the deviations of the angle between the IGs and VCGs trace the shocks in diffuse media. Similarly the perpendicular orientation of the VCGs and the SIGs or to the dust polarimetry data trace the regions of gravitational collapse. We demonstrate the utility of combining the VCGs, IGs and polarimetry using GALFA HI and Planck polarimetry data. We also provide an example of synergy of the VCGs and the SIGs using the HI4PI full-sky HI survey together with the Planck synchrotron intensity data.
\end{abstract}

\section{Introduction}

Interstellar medium is turbulent with magnetic fields playing a critical role in the key processes, including star formation, propagation and acceleration of cosmic rays as well as regulating heat and mass transfer between different phases. Therefore it is essential to have a reliable way of studying magnetic fields. Unfortunately, there only a few ways of tracing magnetic fields and their applicability is limited. For instance, tracing magnetic fields with aligned grains (see \citealt{2007JQSRT.106..225L}) requires either the availability of the background stars, if the polarization via extinction is employed, or sensitive far infrared measurements, which frequently imply the use of the space-based instruments. Another major technique that employs synchrotron polarization mostly traces magnetic fields in the warm and hot phases of the interstellar medium (ISM) (see \citealt{Draine2011PhysicsMedium} for the list of the idealised ISM phases).

Using spectroscopic data for tracing magnetic fields was first suggested and demonstrated in \cite{2002ASPC..276..182L} using the anisotropy of correlation functions of the velocity channel maps. The new approach was motivated by the theoretical advancements establishing that the MHD turbulence is anisotropic (\citealt{GS95}, herafter GS95,  \citealt{Lazarian1999ReconnectionField,Cho2000TheTurbulence,Maron2000SimulationsTurbulence, Kowal2010VelocityScalings}, see \citealt{Brandenburg2013AstrophysicalTurbulence} for a review) and should translate into the anisotropy of the observed correlation fluctuations. Later this approach was employed with the use of velocity centroids \cite{EL05} and \cite{2002ApJ...566..276B} and the Principal Component Analysis (PCA) \cite{2008ApJ...680..420H} of the spectroscopic data. In its PCA incarnation the idea of tracing magnetic field anisotropies was applied to the observational data in \cite{2008ApJ...680..420H} with the results of anisotropies shown to be consistent with the starlight polarization measurements. Further development of the technique resulted in the development of techniques using the anisotropies for obtaining the media magnetization, as it parameterized by the Alfven Mach number \citep{EL05, 2015ApJ...814...77E, 2011ApJ...736...60T} and the separation of compressible and incompressible turbulence components of observational data as discussed in \cite{2016MNRAS.461.1227K,2017MNRAS.464.3617K}.

On the basis of the aforementioned improved understanding of the nature of MHD cascade a new technique employing velocity centroid gradients (VCGs) was introduced in \citeauthor{GL17} (\citeyear{GL17}, henceforth GL17). Instead of appealing to anisotropic velocity correlations as in \cite{2002ASPC..276..182L}  the new technique made use of velocity gradients that, according to the MHD turbulence theory tend to be perpendicular to the {\it local} direction of the magnetic field.\footnote{The turbulent motions in MHD turbulence take place in relation to the {\it local} direction of magnetic field (see \citealt{Lazarian1999ReconnectionField,Cho2000TheTurbulence,Maron2000SimulationsTurbulence}). This fact is crucial for magnetic field tracing with the VCGs and other gradient-based techniques.} This idea was further extended in \citeauthor{YL17} (\citeyear{YL17}, henceforth YL17) where a robust practical procedure for VCG calculations was suggested. In the latter paper we also applied the new technique to the GALFA 21 cm data and demonstrated a good correspondence between the magnetic field tracing with the VCGs and the Planck polarization arising from aligned grains. This result motivates the further study of VCGs properties that we undertake in this paper. 

We note parenthetically that VCGs are not the only way to trace magnetic field using velocity gradients. Intensity gradients within the Position-Position- Velocity (PPV) channel maps have been recently suggested as an alternative way of exploring the magnetic field topology \cite{LY17}. On the basis of the earlier studies of the statistics of channel maps versus the statistics obtained with velocity centroids (see \citealt{EL05,2017MNRAS.464.3617K} and ref. therein) one can expect that the velocity channel gradients in \citeauthor{LY17} (\citeyear{LY17}, henceforth LY17) may have advantages for studying supersonic turbulence as well as tracing the 3D distribution of turbulence. At the same time, the VCGs may have advantages for studying subsonic turbulence. The comparison of the two technique will be done elsewhere. 

Another way of exploring the gradients was introduced in \citeauthor{LYLC17} (\citeyear{LYLC17}, henceforth LYLC17). This approach employs synchrotron intensity gradients (SIGs) and provides a way of tracing magnetic field in turbulent environments the synergy of which to the VCG technique we explore in this paper. The theoretical basis for the SIGs is the same as for the VCGs. However, in the presence of gravity, the response of the SIGs and the VCGs is different which provides a way to identify the regions where the matter is subject to the gravitational collapse, as we discuss in the paper.

The density statistics is also affected by MHD turbulence.  MHD turbulence can imprint some of its properties to density, making its statistics at low sonic Mach numbers anisotropic (see \citealt{Cho2003CompressibleImplications,Beresnyak2005DensityTurbulenceb,2007ApJ...658..423K}). While both from theory and simulations we expect the density to be a worse tracer of magnetic fields\footnote{This was also demonstrated earlier in GL17 and YL17 with numerical simulations.}, especially at high sonic Mach numbers, the very differences between the velocity and anisotropies can be informative of the properties of the turbulent media. This is the synergy that we also explore in this paper. We note that the first report of the correlation of the magnetic fields and density gradients was done within the numerical study by  \cite{Soler2013} reported the alignment of density gradients and magnetic fields with simulation as well as the change of the alignment that occurs after the effects of self-gravity get important starting with a particular density threshold. In our study we make a step further and explore the point-wise tracing of magnetic fields by the intensity gradients (IGs) as well as the nature of the deviations of the IGs directions from those of magnetic field. We show a significant synergy of using the IGs, the VCGs and the dust polarimetry data. In addition, we provide a detailed study on how self-gravity lead to the rotation of both VCG and IG vectors and the physics behind it. We also explore the ways of improving the IG tracing of magnetic field by suggesting the procedure filtering for the shocked gas regions. 

As a separate but related development it is important to mention the study in \cite{Clark2015NeutralForegrounds}. There a good correlation of directions of the filaments observed in the PPV GALFA channel maps and the magnetic field direction as revealed by Planck polarimetry was reported. As a result the filaments were identified as a tracer of magnetic field direction in HI high latitude gas. While in the original study the filaments were identified with the actual spacial density enhancements, we believe that the interpretation depends on the thickness of the velocity channel used. Indeed, according to \cite{LP00} within the sufficiently thin velocity channels the intensity fluctuations are caustics mostly induced by velocity fluctuations. Therefore, depending on the velocity channel thickness, the filament tracing is related either to revealing of velocity statistics or density statistics or both. The comparison of the empirical techniques is an interesting problem that we will discuss in another paper.
   
Self-gravity is expected to modify the properties of the flow in the vicinity centers of the gravitational collapse. The change of the properties of MHD turbulence within such regions is expected to be the result of VCGs changing their properties, in particular their alignment with the magnetic fields. The density distribution is even more affected by self-gravity. Indeed, a recent study by \cite{Soler2013}, reported that the IGs rotate with respect to magnetic fields when arriving some density threshold in super-Alfvenic high density self-gravitating simulation. While the work from \cite{Soler2013} does not address how self-gravity dynamically change the alignment of IGs and VCGs, that encourages a detailed study on how self-gravity can possibly lead to the rotation of gradient vectors and the physics behind it. In this paper, we address the issue of how VCGs and IGs behave in the regions of self-dominant gravity for the case of sub- and trans-Alfvenic clouds, how does the stage of collapse correlated to the alignment of gradients, and how can the aforementioned phenomenon can be used on tracking self-gravitating regions in observation.

In what follows we briefly summarize the theoretical arguments that justify the use of the VCGs in \S \ref{sec:1}, explain our simulation setting in \S \ref{sec:1.5} and numerical approach for gradient calculation in \S \ref{sec:2}. We provide a comparison of the alignment of VCGs and IGs with magnetic fields in different physical conditions in \S \ref{sec:3}, and providing methods on improving the alignment. We explore the properties of the VCGs in sub-Alfvenic self-gravitating clouds in \S \ref{sec:4} and explore their synergistic use with the IGs. We illustrate our method in \S \ref{sec:5} using observational data in \S \ref{sec:5}. We compare our methods to the method of anisotropy of correlation in \S \ref{sec:6}. We compare our method to others and discuss our results in \S \ref{sec:7}. We summarize our results in \S \ref{sec:8}.

\section{Theoretical foundations of the technique}
\label{sec:1}

The theoretical motivation for studying magnetic fields using VCGs are discussed in our earlier papers (GL17, YL17, LY17) as well as in the paper on the SIGs (LYLC17). In short, in the presence of fast turbulent reconnection \citep{Lazarian1999ReconnectionField} the motions that do not induce magnetic field bending are preferentially excited. These are the motions perpendicular to the local direction of magnetic field. This fact is numerically confirmed in \citep{Cho2000TheTurbulence,Maron2000SimulationsTurbulence} and is the corner stone of the modern theory of MHD turbulence (see 
\citealt{Brandenburg2013AstrophysicalTurbulence}). Due to this alignment of velocity motions, the  gradient of velocities are largest perpendicular to the {\it local} direction of magnetic field. It is important that the gradient amplitude for the GS95 turbulence is increasing with the decrease of the eddy scale.\footnote{This is also true for the weak MHD turbulence that takes place for strongly magnetized media with the Alfven velocity much larger than the turbulent velocities (see  \citealt{Lazarian1999ReconnectionField,G2000}). This turbulence transfers to the GS95 turbulence at small scales.} Therefore the smallest eddies that are most aligned with the local direction of the magnetic field contribute most to the gradients. As a result, by tracing the velocity gradients one can trace the magnetic field direction.

The motions at the turbulence injection scale can be super-Alfvenic, i.e. faster than the Alfven velocity $V_A$. The GS95 theory is not applicable to such motions. However, as the turbulent velocity decrease with the scale, e.g. for Kolmogorov turbulence $V_l\sim l^{1/3}$, for a realistically extended turbulence inertial range at small scales the motions gets subAlfvenic and obey the GS95 theory. Therefore, by filtering the contributions from larger scales we may still trace the magnetic fields using velocity gradients.  

In the presence of gravity, the velocity field is modified. In particular, for subcritical collapse when magnetic field dominate the dynamics, the motions are expected along the stiff magnetic field lines, 
making velocity gradients aligned with magnetic field. Similarly, for supercritical collapse, the collapsing material is expected to drag relatively weak magnetic fields. This should also make magnetic fields and velocity gradients aligned. In other words, for the regions with significant self-gravity effects we expect the change of the direction of magnetic field and velocity gradient alignment from being perpendicular to magnetic field to becoming parallel to magnetic field. (See Section \ref{subsec:4.1} for a detailed discussion)

In GL17 and YL17 the velocity centroids were used as a measure of velocity that is available from observations. The velocity centroids were shown to reflect the statistics of underlying velocity field well, at least for subsonic MHD turbulence \citep{EL05}. In the present paper we explore numerically the utility for the VCGs to trace magnetic fields in MHD turbulence for various turbulent driving environments, e.g. different combination of sonic and Alfvenic Mach number, different compression of fluids in non self-gravitating as well as self-gravitating environment. We shall also establish the guidelines on how to get better alignments numerically even the gradient technique does not provide a good prediction.

\section{Simulation}
\label{sec:1.5}

As there are many factors affecting the alignment of rotated gradients with magnetic fields, we would  investigate VCGs and IGs through studying MHD simulations in different physical conditions.  Following YL17 and LYLC17 we used a series of compressible, turbulent, isothermal single fluid magnetohydrodynamic (MHD) simulations for examination of VCGs and IGs from ZEUS-MP/3D \citep{Hayes2006}, a variant of the code ZEUS-MP \citep{Norman2000}. We set up simulation cubes with several combinations of sonic Mach number $M_s$ and Alfvenic Mach number $M_A$, which are listed in Table \ref{tab:simualtionparameters}. For supersonic simulations, we increases the resolution of the simulations to ensure the cube has a sufficiently long inertial range for high $M_s$ cases. The initial magnetic field is along the z-axis and the boundary condition is triply periodic. We injected turbulence using the same method in YL17. To study the effect of compressibility to alignment of gradients, we also introduce a cube (Set E) from an incompressible simulation (See LYLC17 for details).

Most of this work is devoted to the studies of diffuse media with the negligible effects of gravity. In section\S \ref{sec:4} we study the case of strongly magnetized media for which self-gravity is important. There we wait until turbulence inside the cube gets saturated and then switch on the self-gravitating module, which employs a periodic Fast Fourier Transform (FFT) Poisson solver. To trace the effect of gravity with respect to time, we take a number of snapshots with even time intervals before and after the of gravity is switch on. We stop our calculation when any pixel in the system violates the Truelove criterion \citep{Truelove1997}, which informs us when to stop the simulation before having numerical artifacts due to self-gravitating collapse. As self-gravity is a dynamic process, we expect systems having same $M_s$ and $M_A$ but different magnetic criticality, that means a system with gravitational to magnetic energy ratio larger than one, will provide very similar final results. Therefore we here only investigate the systems with magnetic criticality $\Phi = 2 \pi G^{1/2} \rho L/B$ to be 2, which ensures the cube will have physical collapse when self-gravity is activated and before the Truelove criterion is violated.

\section{Numerical Method}
\label{sec:2}

The raw data from simulation cube has to be converted to synthetic maps for our gradient studies. Assuming the line-of-sight direction is the $x$-axis, the intensity $I({\bf r})$, velocity centroid $C({\bf r})$  are defined as
\begin{equation}
\begin{aligned}
I({\bf r}) &= \int \rho({\bf r},x)  dx\\
C({\bf r}) &=I^{-1} \int \rho({\bf r},x) v_x({\bf r},x)  dx
\end{aligned}
\end{equation}

We also computed polarization by assuming the media is optically thin, Faraday rotation is not effective along the line of sight, and the emissitivity is a constant over the line of sight. In other words, the Stokes parameters $I_s({\bf r}), Q({\bf r}), U({\bf r})$ can be expressed in terms of $B_{y,z}$,  the y and z direction magnetic fields: 
\begin{equation}
\begin{aligned}
I_s({\bf r}) &\propto \int B_z^2+B_y^2  dx\\
Q({\bf r}) &\propto \int B_z^2-B_y^2  dx\\
U({\bf r}) &\propto \int -2B_yB_z  dx
\end{aligned}
\end{equation}
The polarized intensity $P=\sqrt{Q^2+U^2}$ and angle $\phi =0.5\tan^{-1}(U/Q)$ are then defined correspondingly. (See \citealt{2016ApJ...831...77L} for details). 

The three observables $I,C,I_s$ will be used to calculate the intensity gradients (IGs), velocity centroid gradients (VCGs) and synchrotron intensity gradients (SIGs), by following the recipe and the {\it sub-block averaging}. The {\it sub-block averaging} method, as it introduced in YL17 is based on defining the most probable direction within a block by a local Gaussian-fitting peak on the distribution of gradient angle distribution. Within the approach, the fitting error provides a quantitative estimate of whether the block size is large enough for providing a correct direction. 

\section{Alignments between VCGs/IGs and magnetic field}
\label{sec:3}
To compare the performance between VCGs and IGs, we employed the {\it alignment measure} that is introduced in analogy with the grain alignment studies (see \citealt{2007JQSRT.106..225L}):
\begin{equation}
AM=2\langle\cos^2\theta_r\rangle-1,
\end{equation}
(see GL17, YL17) with a range of $[-1,1]$ to measure the relative alignment between {\it rotated} gradients and magnetic fields, where $\theta_r$ is the relative angle between {\it rotated} gradients to magnetic field.  A perfect alignment gives $AM=1$. On the other hand, $AM=-1$ when vectors tend to be more perpendicular. We shall use AM to quantify the performance of the VCGs and IGs in various physical conditions. Some special values for AM can also give the readers how good the alignment it is for certain type of gradient vectors to magnetic fields. For example, $AM=0.5$ indicates the dispersion between {\it average} gradients to magnetic field directions is about $15^o$, which means the gradient vectors are rather good tracers for the magnetic field direction. On the other hand, $AM=0$ indicates a random orientation of gradient vectors relative to magnetic fields. In what follows, we shall study both VCGs and IGs and investigate their advantages and synergy for studying magnetic field and other parameters of the ISM.  

\subsection{Identifying shocks with gradients}
\label{subsec:3.1}

\begin{figure}[t]
\centering
\includegraphics[width=0.44\textwidth]{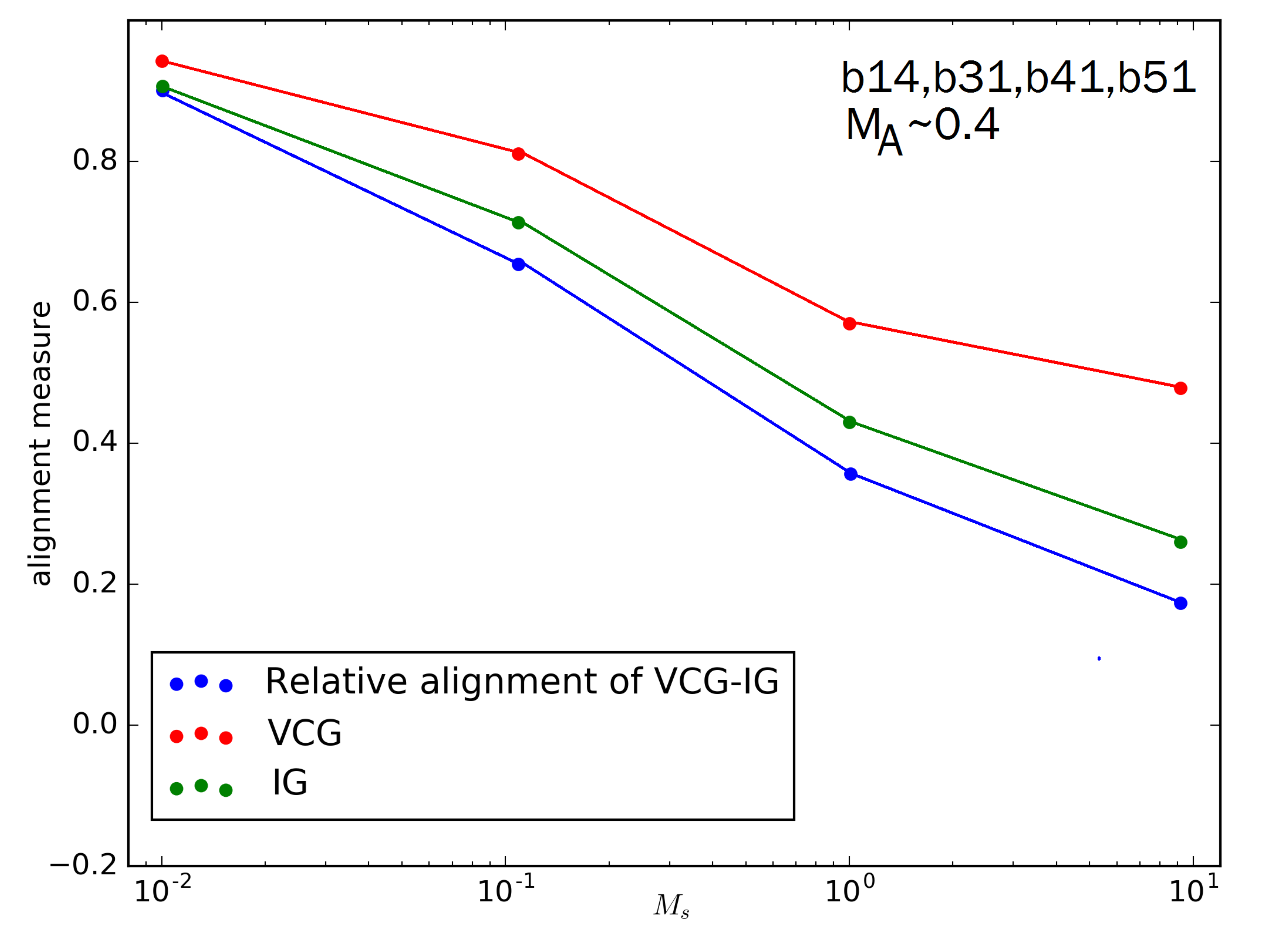}
\caption{\label{fig:VCGIG-Ms} A diagram showing the alignment of VCGs (red), IGs (green), and the relative alignment between them (blue) as the sonic Mach number $M_s$ increases}
\end{figure}

Shocks are a very important process when the sonic mach number, $M_s$, which is the ratio of the turbulent injection velocity and the speed of sound, gets large. Different phases of the ISM are expected to have different sonic Mach numbers $M_s$ (see \citealt{Burkhart2009}). The Mach number $M_s$ of warm gas is of the order of unity (see \citealt{Gaensler2011Low-Mach-numberGradients} and ref. therein), while the Mach number of cold interstellar media (see \citealt{2009ASPC..414..453D}) as well as molecular clouds can be $10$ and even higher. The sonic Mach number also characterize the compressibility of turbulent flow. When $M_s$ increases, stronger compressions result in stronger shocks and significantly modify the properties of turbulent gas. We study how shocks change the alignment of gradient vectors and the underlying magnetic field. We start our investigation with the cubes from set "B" (see Table \ref{tab:simulationparameters}) to perform an alignment test on both the VCGs and the IGs. Figure \ref{fig:VCGIG-Ms} shows how the VCGs and the IGs behave with various $M_s$. We note that the alignment of the VCGs in significantly better compared with the IGs.  

\begin{table}
 \centering
 \label{tab:simualtionparameters}
 \caption {Simulations used in our current work. The magnetic criticality $\Phi$ is set to be 2 for all simulation data.}
 \begin{tabular}{c c c c c}
  Set & Name & $M_s$ & $M_A$ & Resolution\\ \hline \hline
  A & a1 & 0.2 & 0.02 & $480^3$\\
  & a2 & 0.2 & 0.066 & $480^3$\\
  & a3 & 0.2 & 0.2 & $480^3$\\
  & a4 & 0.2 & 0.66 & $480^3$\\
  & a5 & 0.2 & 2.0 & $480^3$\\
  & a6 & 0.66 & 0.2 & $480^3$\\
  & a7 & 0.02 & 0.2 & $480^3$\\
  B & b11 & 0.4 & 0.04 & $480^3$\\
  & b12 & 0.8 & 0.08 & $480^3$\\
  & b13 & 1.6 & 0.16 & $480^3$\\
  & b14 & 3.2 & 0.32 & $480^3$\\
  & b15 & 6.4 & 0.64 & $480^3$\\
  & b21 & 0.4 & 0.132 & $480^3$\\
  & b22 & 0.8 & 0.264 & $480^3$\\
  & b23 & 1.6 & 0.528 & $480^3$\\
  & b31 & 0.4 & 0.4 & $480^3$\\
  & b32 & 0.8 & 0.8 & $480^3$\\
  & b41 & 0.132 & 0.4 & $480^3$\\
  & b42 & 0.264 & 0.8 & $480^3$\\
  & b51 & 0.04 & 0.4 & $480^3$\\
  & b52 & 0.08 & 0.8 & $480^3$\\
  C& c1 & 5 & 0.2 & $792^3$\\
  & c2 & 5 & 0.4 & $792^3$\\
  & c3 & 5 & 0.6 & $792^3$\\
  & c4 & 5 & 0.8 & $792^3$\\
  & c5 & 5 & 1.0 & $792^3$\\
  & c6 & 5 & 1.2 & $792^3$\\
  & c7 & 5 & 1.4 & $792^3$\\
  & c8 & 5 & 1.6 & $792^3$\\
  & c9 & 5 & 1.8 & $792^3$\\
  & c10 & 5 & 2.0 & $792^3$\\
  D & d10 & 5 & 2.0 & $1200^3$\\
  E (Incompressible) & e & 0 & 0.8 & $512^3 $ \\ \hline
 \end{tabular}
\end{table}

Supersonic turbulence will create strong density compression associated with shocks, which, as a result, create compressional modes rather than the incompressible Alfvenic modes. In contrast, the approach underlying magnetic field tracing by velocity gradient is justified for the incompressible Alfvenic modes (GS95, \citealt{Cho2003CompressibleImplications}), for which the eddies are elongated along the local magnetic field direction, causing the velocity gradients to be maximally perpendicular to the local magnetic field direction. In fact, the fast mode gradients tend to be more isotropic compared to that of slow and Alfvenic modes, which both of them are maximal perpendicular to local magnetic field directions.Therefore a higher sonic Mach number should reduce the alignment, as illustrated in Figure \ref{fig:VCGIG-Ms}. As a reference, the average ratio between modes are approximately 0.7:0.2:0.1 in the order of Alfven, slow and fast modes. However, it is in general not true that compressibiltiy will decrease alignment, as the slow mode resulted from compression still contributes like that of Alfvenic mode in terms of gradients \cite{LY17}. 

\begin{figure}[t]
\centering
\includegraphics[width=0.44\textwidth]{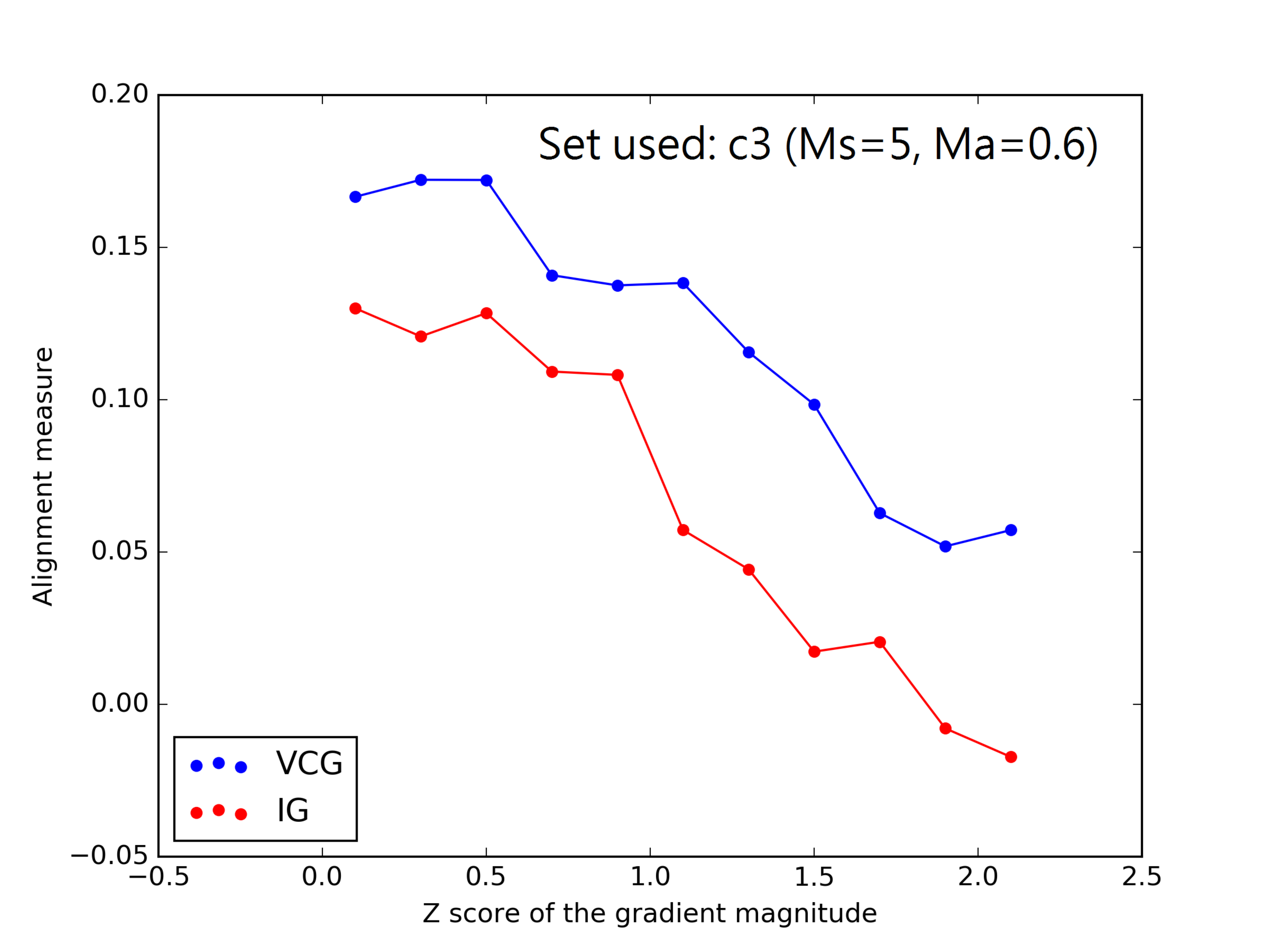}
\caption{\label{fig:2} A plot showing how AM of VCGs and IGs depend on the IGs amplitude.}
\end{figure}

\begin{figure*}[t]
\centering
\includegraphics[width=0.98\textwidth]{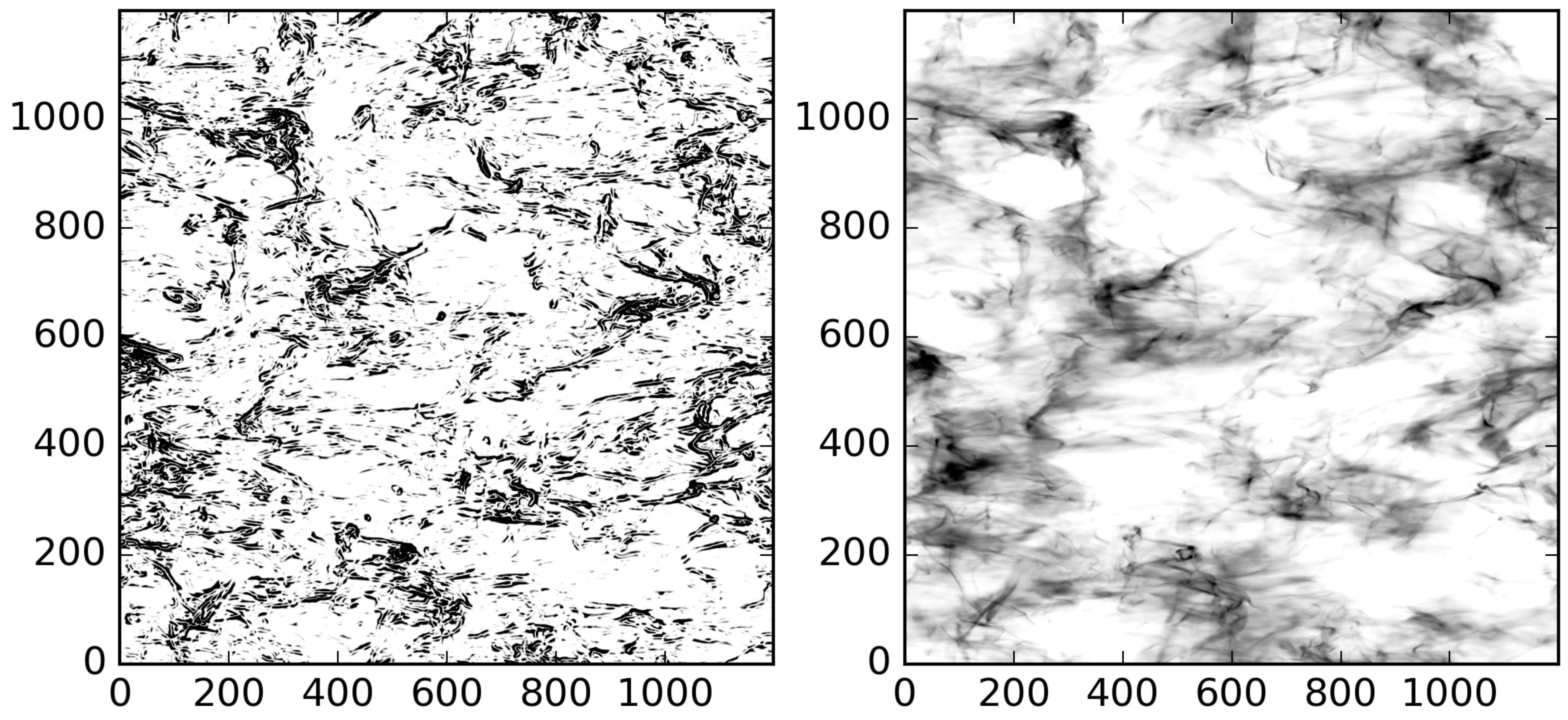}
\caption{\label{fig:shock} Two figures showing how the intensity gradient amplitude (left) is related to high density shock structure (right). }
\end{figure*}

We note that the VCGs trace magnetic fields better that the IGs in supersonic turbulence. The reason is that, as we mentioned earlier, density statistics is an indirect tracer of magnetic fields \citep{Beresnyak2005DensityTurbulenceb,2007ApJ...658..423K},  but they it is directly influenced by shocks. To improve the IG alignment in high-$M_s$ environments, we apply a shock removal algorithm to our synthetic maps. To start with, we focus on removing the strong J-shocks (see \citealt{2009ASPC..414..453D}). For Jump discontinuity, the change of density across the shock is very significant compared to the surrounding environment. Therefore a higher density gradient amplitude is found across the shock front. Hence we sort out both the VCGs and IGs data according to IGs amplitude, and check the alignment on the slice of  gradient data using a $M_s=5$ map in Figure \ref{fig:2} for both the VCGs and the IGs.  Given the IGs amplitude $x$, and the global mean $\mu$, the standard deviation $\sigma$, the Z-score of x is defined as $Z(x)=(x-\mu)/\sigma$. A higher positive Z-score stands for regions with gradient amplitude above the system average. Regions with higher amplitude correspond to those with J-shocks. To focus on the high IGs amplitude regions, we only plot regions with positive Z-scores. Figure \ref{fig:shock} shows how the shock identification algorithm works on the high resolution data D10. The intensity gradient map is in the form of Z-score, and only the positive Z-score part is shown. From inspection, there is indeed one-one correspondence between the shock boundary and the shock structure. The promising result from Figure \ref{fig:shock} encourages our study of Z-score to alignment below.

Figure \ref{fig:2} shows that higher gradient amplitude tend to have weaker alignment. For simplicity, we compute the alignment without applying the sub-block averaging. That means that the data points in \ref{fig:2} are being classified only by their respective gradient amplitudes. A very important message from Figure \ref{fig:2} is the alignment will decrease when getting close to the shock front. We expect that removing pixels with higher gradient amplitudes could help improving the performance of gradient technique globally. At the same time, the pixels that we remove for the gradients to work better carry the information about the shocks and such regions can be studied separately.

We do, however, have to emphasize that that this is only a preliminary study of the effects of the shocks and their removal. One particular question that has to address is the distinct behavior of the three MHD modes in systems with different $M_s$ and $M_A$, which is the more fundamental milestone on study magnetized turbulence (see \citealt{Cho2003CompressibleImplications}), gives drastically different alignment in terms of VCGs. Nevertheless, shocks in strongly magnetized media are usually align perpendicular or parallel to magnetic field, depending $M_s$ and $M_A$ and the ratio of modes. We discuss the effects of different modes in our work on the velocity channel gradients \cite{LY17}. 

\subsection{How does Alfvenic Mach number $M_A$ affect gradient alignment?}
\label{subsec:3.2}

Aside from $M_s$, the Alfvenic Mach number $M_A$, which is the ratio of the turbulent velocity to the Alfven velocity, also characterize of the properties of magnetized turbulence. For large $M_A$ turbulent motions are not affected by magnetic fields and tend to be more isotropic.  The turbulent cascade is different when $M_A$ is less than 1 or larger than 1 (See {\citealt{Lazarian1999ReconnectionField,EL05}). 
When $M_A<1$, $k_{||} \propto k_{\perp}^{2/3}$ \citep{GS95}. On the other hand, for $M_A>1$ the nature of the cascade changes at a particular scale $l_A=LM_A^{-3}$, where L is the injection scale
\citep{Lazarian2006EnhancementTurbulence}. When $1/l_A<1/k<L$, the motions are of a hydrodynamic nature, while for smaller scales the GS95 cascading takes over. Therefore we expect that our reasoning based on the GS95 theory should be modified for this range of spacial scales.   

Below we test the alignment in simulations with various $M_A$.  In this test, we use the set C in Table \ref{tab:simualtionparameters}. Figure \ref{fig:3} shows the morphology of VCGs (rotated 90 degrees) as well as the projected magnetic field. For trans-Alfvenic systems in the lower panels, the rotated velocity gradient vectors trace magnetic fields reasonably well, which is consistent with our earlier results in YL17 and GL17. 
On the other hand, as expected, gradients in super-Alfvenic turbulence behave differently. The motion scales larger than $l_A$ are hydrodynamic (see \citealt{Lazarian2006EnhancementTurbulence}) and thus the alignment of gradients at these scales is different from that related to subAlfvenic motions. To mitigate the effect of super-Alfvenic motions, we introduce a low wavenumber filtering: we remove Fourier modes with the wavenumber below a certain threshold. We know from LYLC17 that removing low spatial frequencies does not much affect the tracing of magnetic fields by gradients and therefore we can expect that magnetic fields can be traced by the VCGs in the super-Alfvenic turbulence where $M_A$ is not too high. (See the discussion on highly super-Alfvenic turbulence in \S \ref{subsec:superalf})

\begin{figure}[h]
\centering
\includegraphics[width=0.44\textwidth]{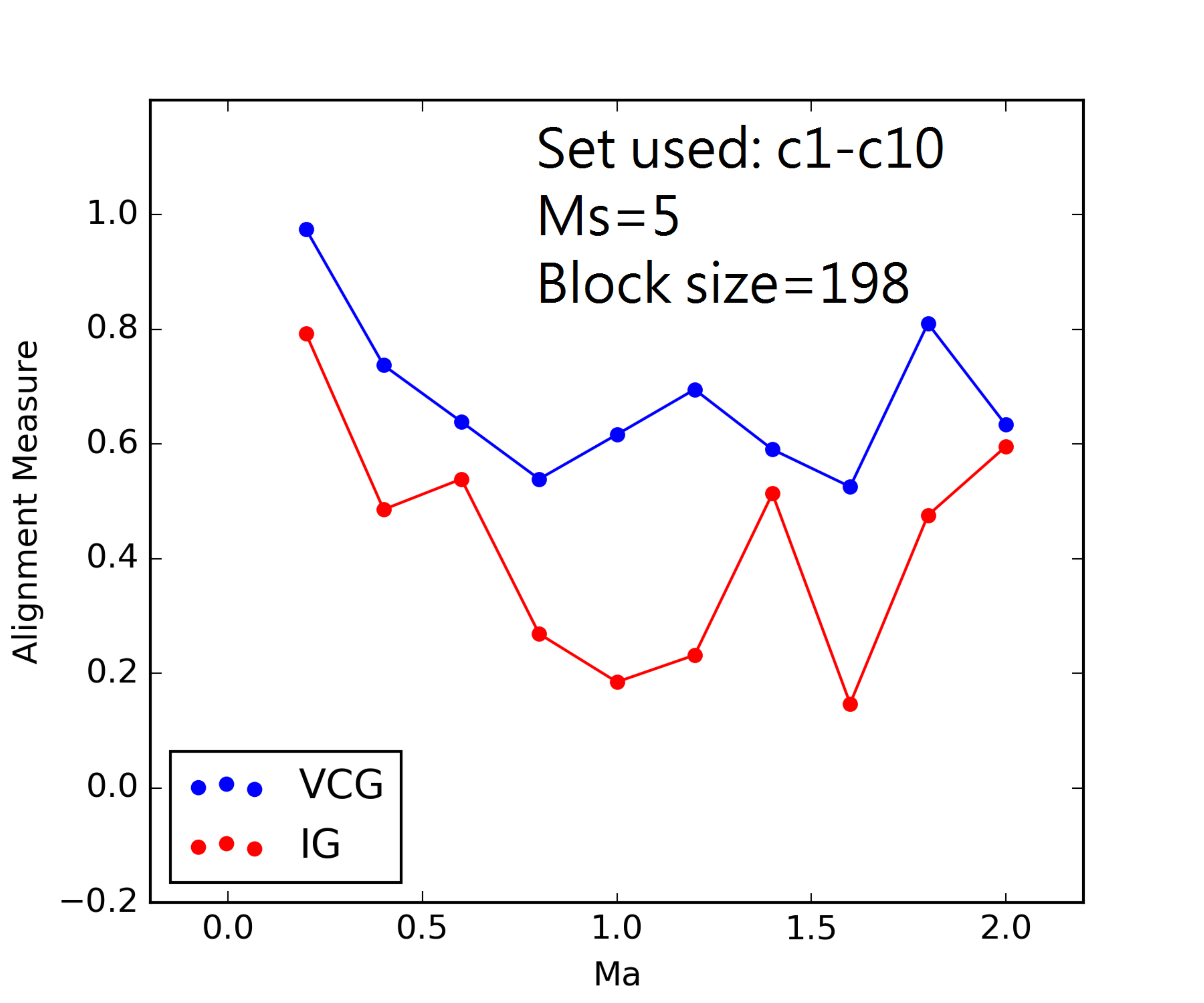}
\caption{\label{fig:3} A plot showing the alignment of VCGs and IGs in different $M_A$.}
\end{figure}

\begin{figure}[h]
\centering
\includegraphics[width=0.44\textwidth]{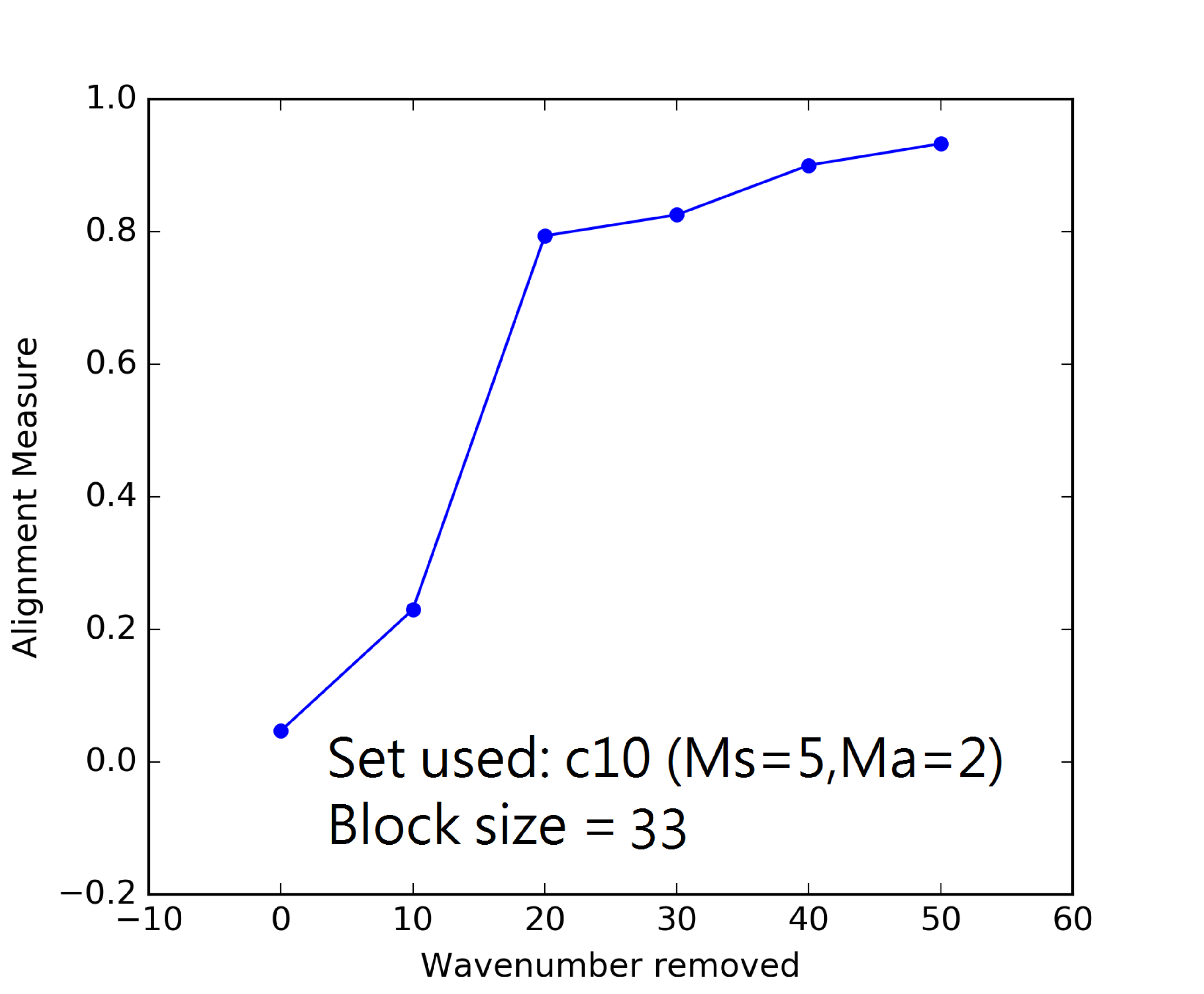}
\caption{\label{fig:4} A plot showing how the wavemode removal algorithm works.}
\end{figure}

We computed the VCGs map obtained after filtering low spatial frequencies in the direction of the projected magnetic field. Figure \ref{fig:4} shows the result of the alignment when wavenumbers corresponding to the scales larger than, respectively, $10,20,30,40$ and $50$ pixels are being removed from a selected cube with resolution $792^3$. We observe that as we remove the contribution from lower wavenumbers, the alignment between the VCGs and projected magnetic fields is gradually increasing. This result matches with our theoretical expectations, as we remove the contribution of the hydrodynamic-type cascade (see \citealt{Lazarian2006EnhancementTurbulence}).  

The practical removal of low-k contributions requires high resolution numerical simulations.  (See \S \ref{subsec:superalf} for a detailed discussion). The underlying reason is, as $M_A$ increases, the low-pass filter value also increases accordingly. With limited inertial range from our simulations, the hydrodynamic-like turbulence proceeds up to the smallest scales affected by the numerical dissipation. Therefore we may not get any expected alignment of the VCGs with the magnetic field. 

Since the anisotropy exists for large scales in sub-Alfvenic systems, the removal algorithm does not change the alignment when facing strong field cases. The corresponding calculations have been performed for the synchrotron intensity gradients (SIGs) in LYLC. The same conclusion from our  previous study as well as the present one indicates that it is advisable to remove the small wavenumber contributions, i.e. produce the low-k filtering of their data, before applying the gradient technique. Elsewhere we will explore the optimization of filtering procedures.

\subsection{How does line of sights in the $B$-direction affect gradient alignment?}

In the general case it is natural to assume that the mean magnetic field has two components perpendicular to the line of sight and a non-zero component parallel to the line of sight. As we discussed earlier, in strong field scenarios, fluid motions are different along and perpendicular to the magnetic field, i.e. the eddies are elongated along the magnetic field. The motions look different when viewed from an oblique angle. To study this effect we perform calculations of the VCGs and the IGs changing the angle between the line of sight and the mean magnetic field. 

We introduced a rotation-projection algorithm (RPA) to mimic the projected maps when the cube is being viewed from different directions. The idea of RPA is to get the rotated coordinates of the rotated three-dimensional cube through oblique projection on a predefined discrete rectangular coordinate plane. For each pixel on the projected plane, it accumulates the data from the three dimensional data with the same integral plane-of-sky coordinates, this mimics the line-of-slight observation. As the number of samples accumulated in the projected plane pixels are different, we perform a simple weighting of the contributions along the line according to the number of samples.  Notice that the periodicity of the simulation domain enables us to perform such an action without worrying about the boundary conditions. To smooth the effect from the rectangular grids, we apply a Gaussian pre-filter of $\sigma=2$ pixels, similar to that that we employed LYLC17 to mitigate the effect of noise. Figure \ref{fig:5} shows the change of AM with respect to the projection angle. We observe that the $AM$ of the IGs is stronger affected by the line-of-sight effect compared to the VCGs. We also notice that  the alignment of the VCGs has a more regular behavior, with the $AM$ gradually decreasing as the angle of rotation approaches $90^o$. 

\begin{figure}[h]
\centering
\includegraphics[width=0.44\textwidth]{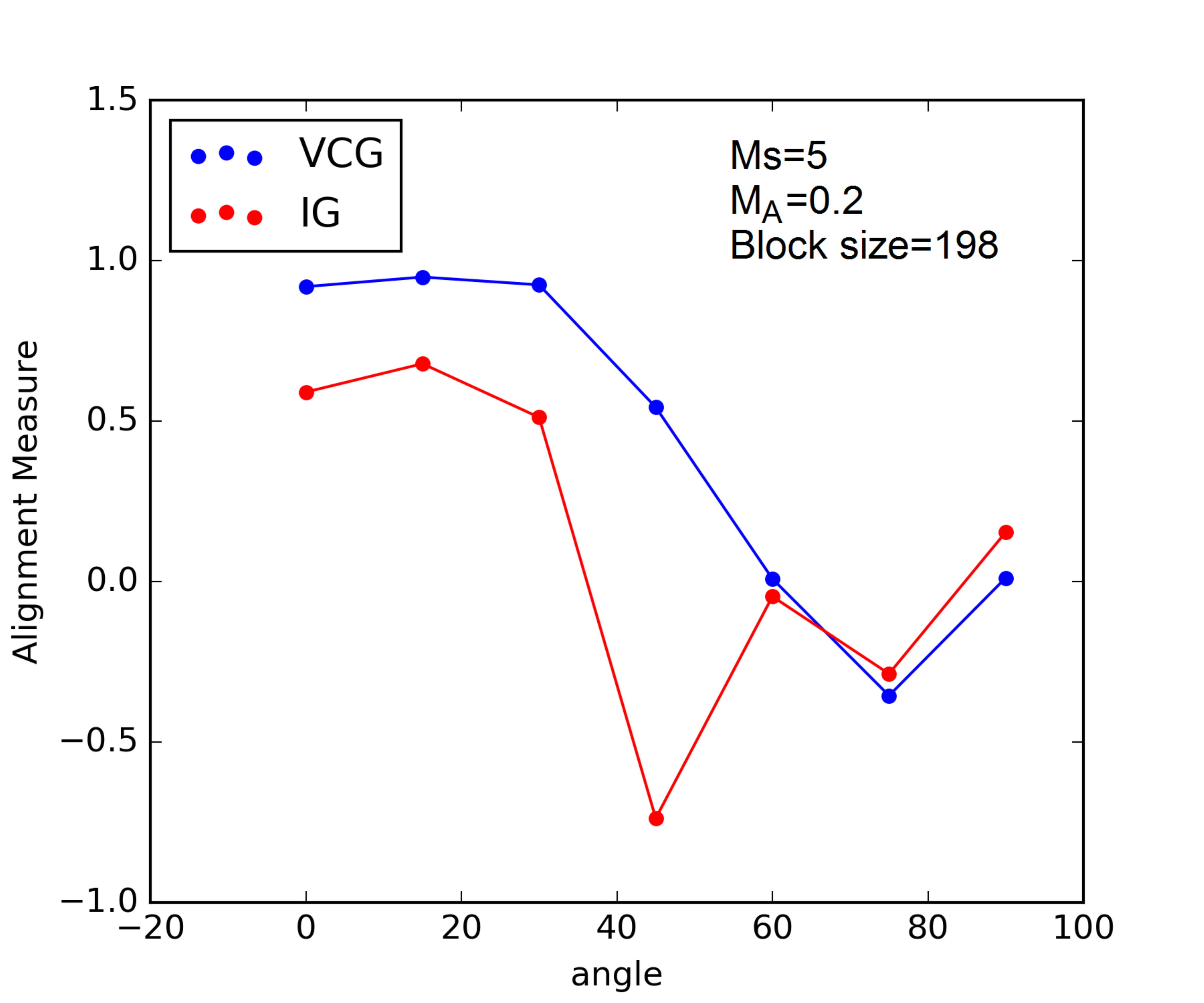}
\caption{\label{fig:5} The $AM$ of VCGs (Blue) and IGs (Red) with respect to the angle of rotation. Defining the line of sight when it is perpendicular to magnetic field to be $0^o$ rotation, and parallel as $90^o$, we compute AM in the projected map.}
\end{figure}

\section{VCGs in self-gravitating regions}
\label{sec:4}

Our study above is applicable to the diffuse gas where MHD turbulence is not modified by self-gravity effects. In molecular clouds we expect the effects of self-gravity to be important.  We expect that both velocity and density statistics respond to self-gravity but their response can be different. Below we provide the first study of self-gravity on the VCGs and this study is limited by the case of dynamically important magnetic fields. 

\subsection{Expectations and numerical resuls}

\label{subsec:4.1}
Self-gravity changes the motion of magnetized turbulent fluid. An additional acceleration induced by gravity to alter the turbulence anisotropy. The relative importance of gravity to magnetic field, which is measured by the magnetic criticality $\Phi\propto \sqrt{{F_{frav}}/{F_{mag}}}$, determines the properties of the turbulence. As a result, velocity gradients carry information about systems with different $\Phi$. Below we explain the change of the VCGs induced by gravity. In parallel, we discuss what is expected for the IGs.

We assume the gravitational center is located at the center of an eddy as depicted in Figure \ref{fig:Gravity-illustration}. The velocity contour, which is equivalent to the streamlines of {\it $90^o$ rotated} velocity gradients, is the densest perpendicular to the local magnetic field direction. The inward acceleration is governed by the density distribution of matter. For the dynamically important magnetic field we expect that in the direction perpendicular to the magnetic field, the gravitational pull inducing the acceleration is counteracted by a magnetic force.  Hence, the gravitational pull induces the largest acceleration of plasma in the direction parallel to the magnetic field. This acceleration causes the gradients of velocities to turn parallel to the magnetic field.  

\begin{figure}[h]
\centering
\includegraphics[width=0.48\textwidth]{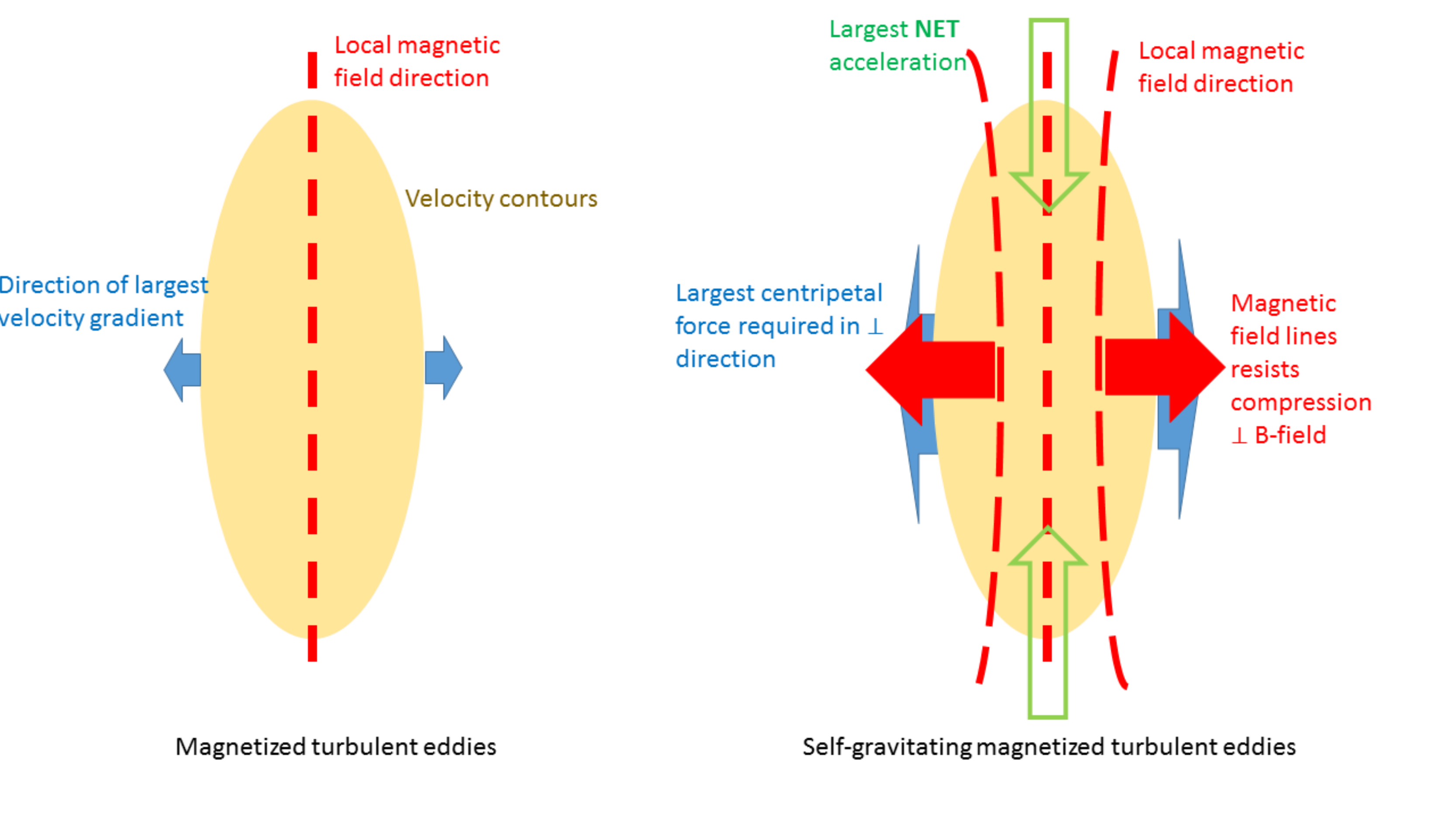} 
\caption{\label{fig:Gravity-illustration} Illustrations on how self-gravity changes the maximum gradient direction. When gravity is absent (Left), eddies are elongated parallel to the local magnetic field direction. Assuming that the gravity center is at the center of the contours we expect a change in the velocity field (Right). }
\end{figure}

\begin{figure*}[t]
\centering
\includegraphics[width=0.98\textwidth]{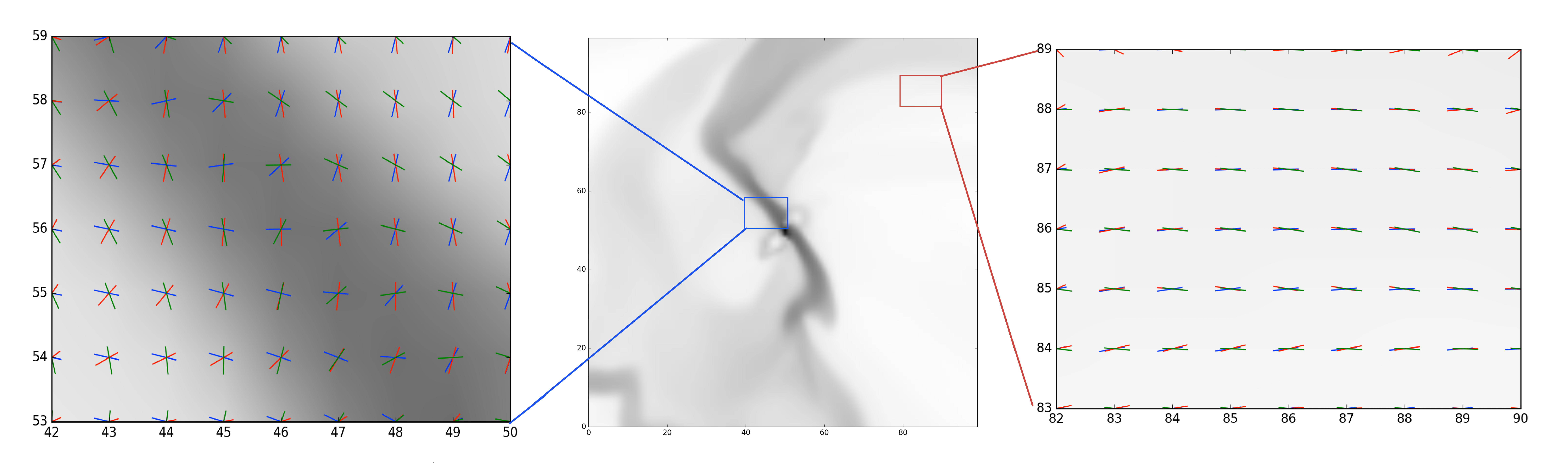}
\caption{\label{fig:grav-dvb}The morphology of the {\it $90^o$ rotated} VCGs (Red), {\it $90^o$ rotated} the IGs (Green) and magnetic field (Blue) on an intensity map, with a gravitating core in the center of the map. We select a 40x40 pixels snapshot for this core. The subpanels illustrate that the gradients behave differently at different distance from the core.}
\end{figure*}

\begin{figure}[t]
\centering
\includegraphics[width=0.45\textwidth]{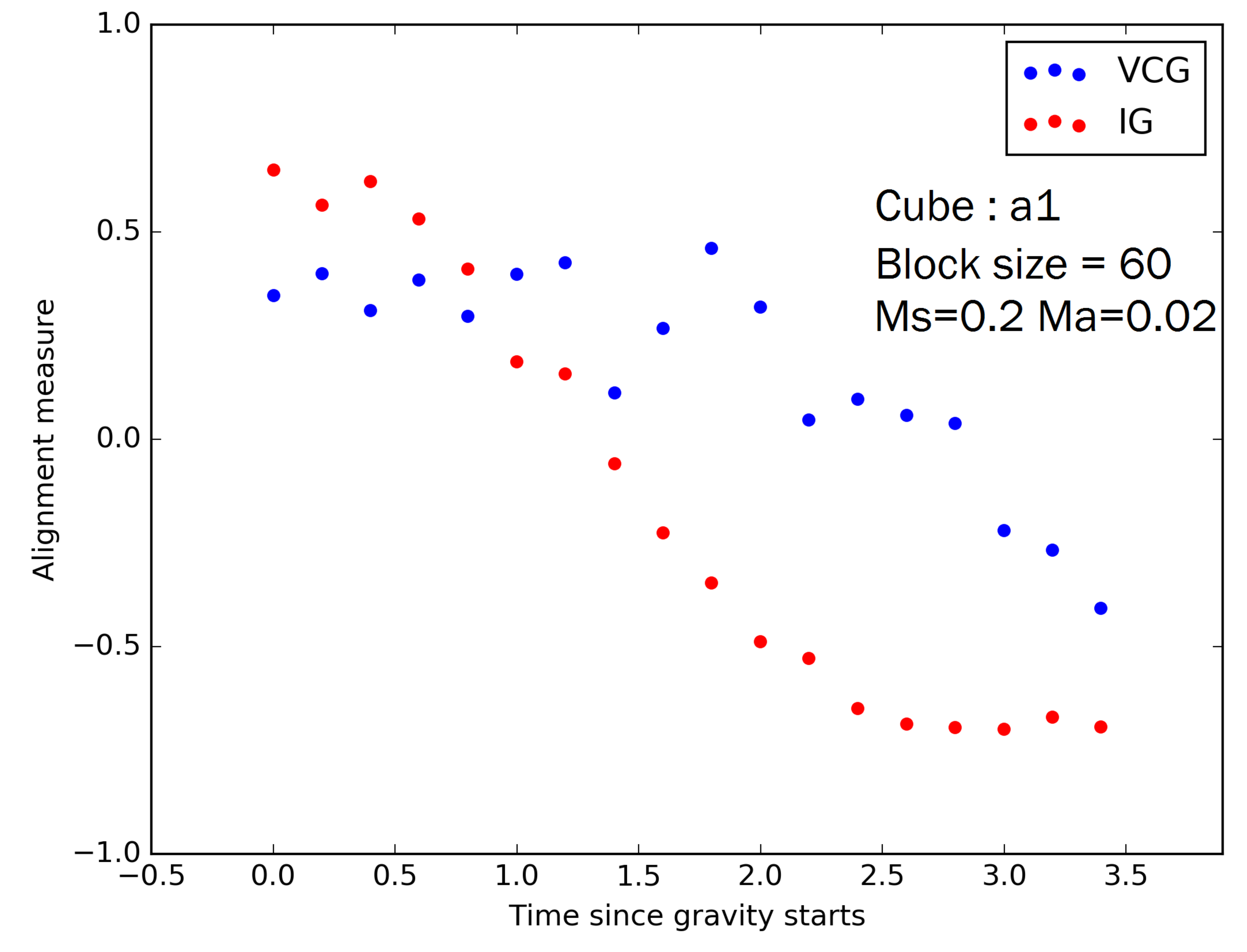} 
\caption{\label{fig:Gravity-AM} Response of AM for VCGs and IGs when gravity is introduced at $t=0$. }
\end{figure}

\begin{figure}[h]
\centering
\includegraphics[width=0.48\textwidth]{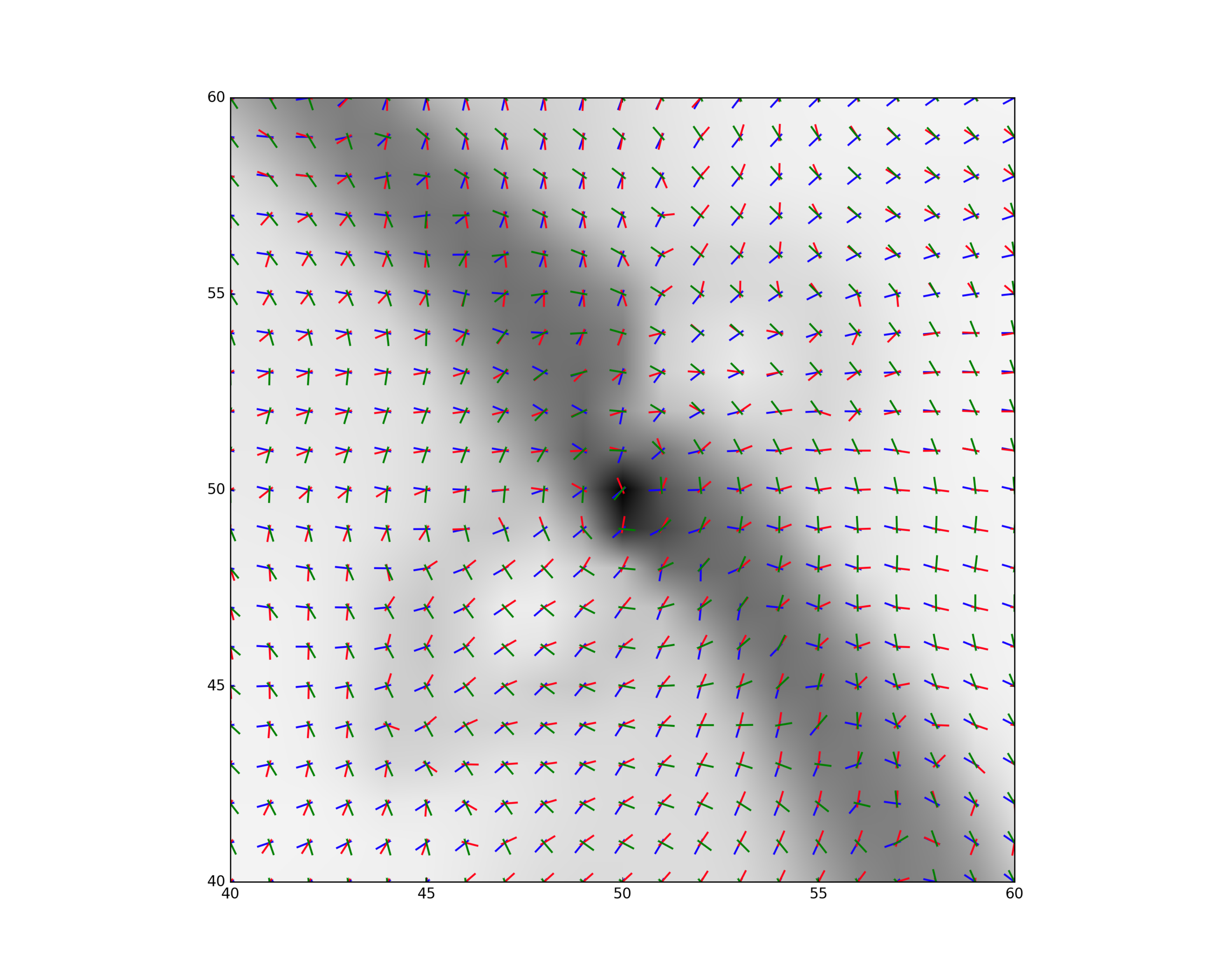}
\caption{\label{fig:grav-zoomin} Similar to Figure \ref{fig:grav-dvb}, but a 20x20 pixels snapshot for this core. The density gradients are almost always perpendicular to the magnetic field. The VCGs and IGs are rotated as in YL17, just presuming themselves tracing magnetic field in diffuse regions. Both {\it rotated} VCGs (Red) and IGs (Green) show significant deviations from the magnetic field direction (Blue). }
\end{figure}

In terms of the VCGs  near the gravity center, the direction of velocity gradients gets {\it parallel} to magnetic field instead of being {\it perpendicular} to magnetic field. On the other hand, the rotation for the IGs depends on the shape of the self-gravitating core during the collapse. If the flat structure perpendicular to the magnetic field is formed, we expect the IGs to also be eventually parallel to the magnetic field. At the same time, in regions away from the gravitational center, the velocity gradient should still be perpendicular to local magnetic field, similar to the case of diffuse media that we studied earlier. These two effects are shown in the left and right panels of Figure \ref{fig:grav-dvb}. 

In the language of the alignment measure $AM$, since the gradients near the gravitational center are turned by $90^o$, the $AM$ of the block covering the core should gradually drop below zero. We tested this conclusion numerically and our results are shown our in Figure \ref{fig:Gravity-AM}. At $t=0$, the gravity is introduced into the cube that already stirred with magnetized turbulence. The collapse of the high density structure starts until some parts of the cube violates the Truelove criterion \citep{Truelove1997}, which happens at the time of $t=3.5$. At $t=1.5$, the AM of IGs falls below zero, while that for VCGs happens at $t\sim 2.8$. After which the negative AM indicates most of the gradient vectors (both the VCGs and the IGs) around the density peak tend to {\it align parallel} to the magnetic field direction. This is also illustrated in \ref{fig:grav-zoomin}.  

Figure \ref{fig:Gravity-AM} also illustrates another important difference in the behavior of the IGs and the VCGs: the AM of IGs changes significantly faster than that of the VCGs. A faster change in the direction of the IGs compared to the VCGs (Table \ref{tab:gradtable}) can be attributed to the fact that the density response directly to gravitational pull, while the direction of the VCGs changes only when the velocity of the gravitational collapse exceeds the turbulent velocities. Since in diffuse media we expect an alignment of IGs and VCGs, the aforementioned effect can actually be observed to identify the regions of gravitational collapse even without using the polarization data:  A gradual change of the relative alignment between the VCGs and the IGs together with the observed increase of intensity with the reversal to the alignment at larger intensities can indicate the onset of gravitational collapse.   

\subsection{Locating self-gravitating regions using VCGs and synchrotron intensity gradients}

\begin{figure}[t]
\centering
\includegraphics[width=0.45\textwidth]{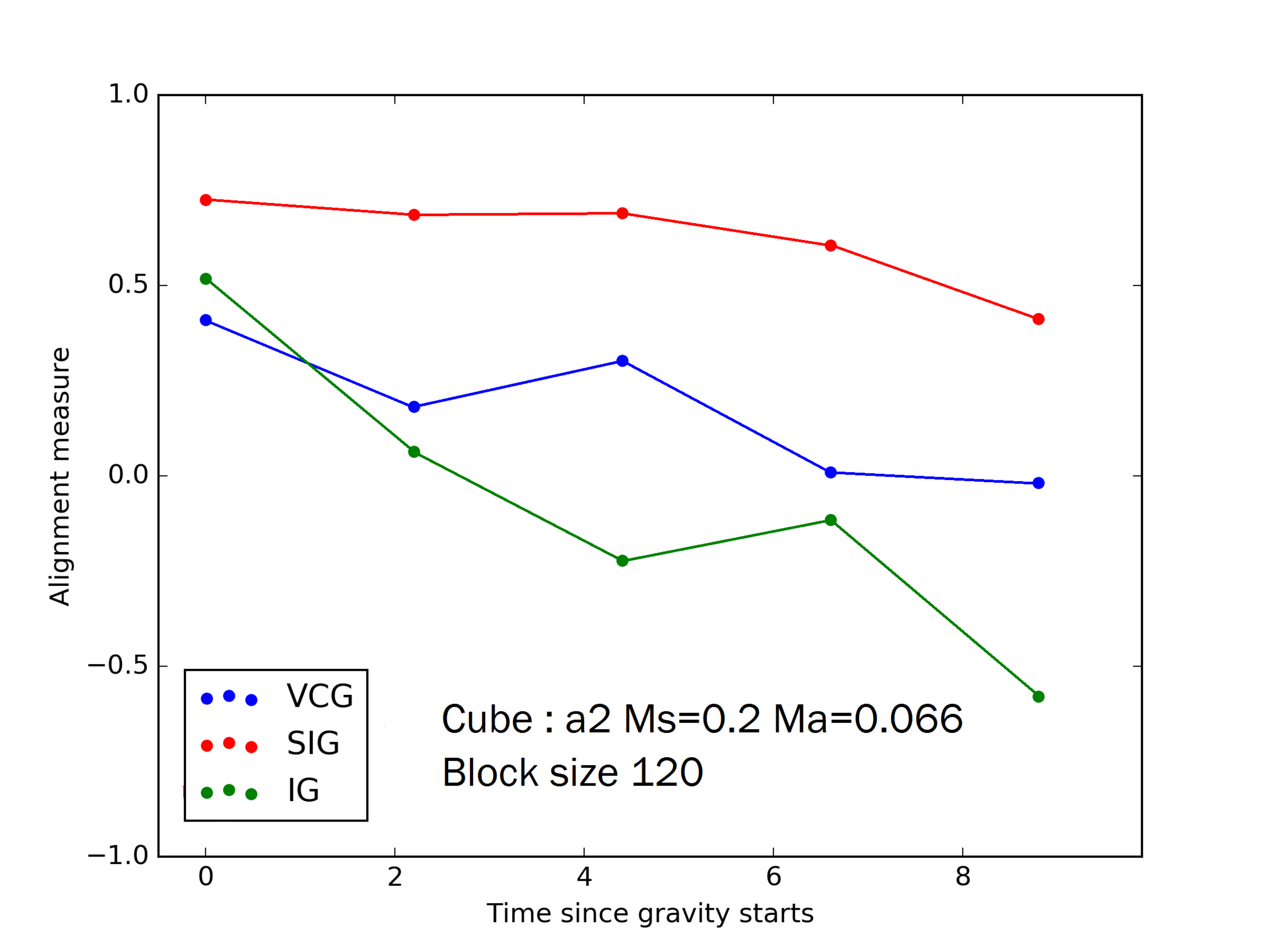} 
\caption{\label{fig:SIG-trend} Response of AM for SIG, VCGs and IGs when gravity is introduced at $t=0$.}
\end{figure}

It is important to determine which parts of the ISM are self-gravitating. Such information can be obtained by comparing gradient vectors from high column density pixels to magnetic field directions, e.g inferred from polarization data. Here we discuss the synergy of the VCGs and SIGs to be used for this purpose.

It was shown in \cite{LYLC17} that SIGs are not expected to change directions when self-gravity becomes important. Therefore when comparing with the 90 degree rotated VCGs, the SIGs in the self-gravitating regions are expected to be perpendicular to them in the regions of gravitational collapse, but parallel in regions with marginal self-gravity effects (Table \ref{tab:gradtable}). We test this expectation by calculating the SIGs and the VCGs as well as the IGs. Figure \ref{fig:SIG-trend} illustrates the difference of the behavior of the AM of SIGs and that of VCGs and IG. The SIGs stays aligned to magnetic field with the AM of $\sim 0.7$ which is changes only marginally. On the other hand, VCGs and IGs both show a decreasing trend as gravity acts, which are shown both in Figure \ref{fig:Gravity-AM} and Figure \ref{fig:SIG-trend}. That suggests that, unlike the SIGs, both theVCGs and the IGs are highly related to the effects of self-gravity.

\begin{table}
 \centering
 \label{tab:gradtable}
 \begin{tabular}{c c c c}
  Types & Diffuse media & Self-gravitating region & Re-rotation speed\\ \hline \hline
  VCGs & $\bot$ & $\parallel$ & Slow\\
  IGs& $\bot$ & $\parallel$ & Fast\\
  SIGs& $\parallel$ &  $\parallel$ & Not rotating \\\hline
 \end{tabular}
  \caption {A summary on the expected direction relative to magnetic field of different class of gradients from GL17,YL17,LYLC17 and this work.}
\end{table}

\subsection{Testing the existence of a density threshold for gradient rotation around the self-gravitating center}
\label{subsec:dthres}
\begin{figure*}[t]
\centering
\includegraphics[width=0.96\textwidth]{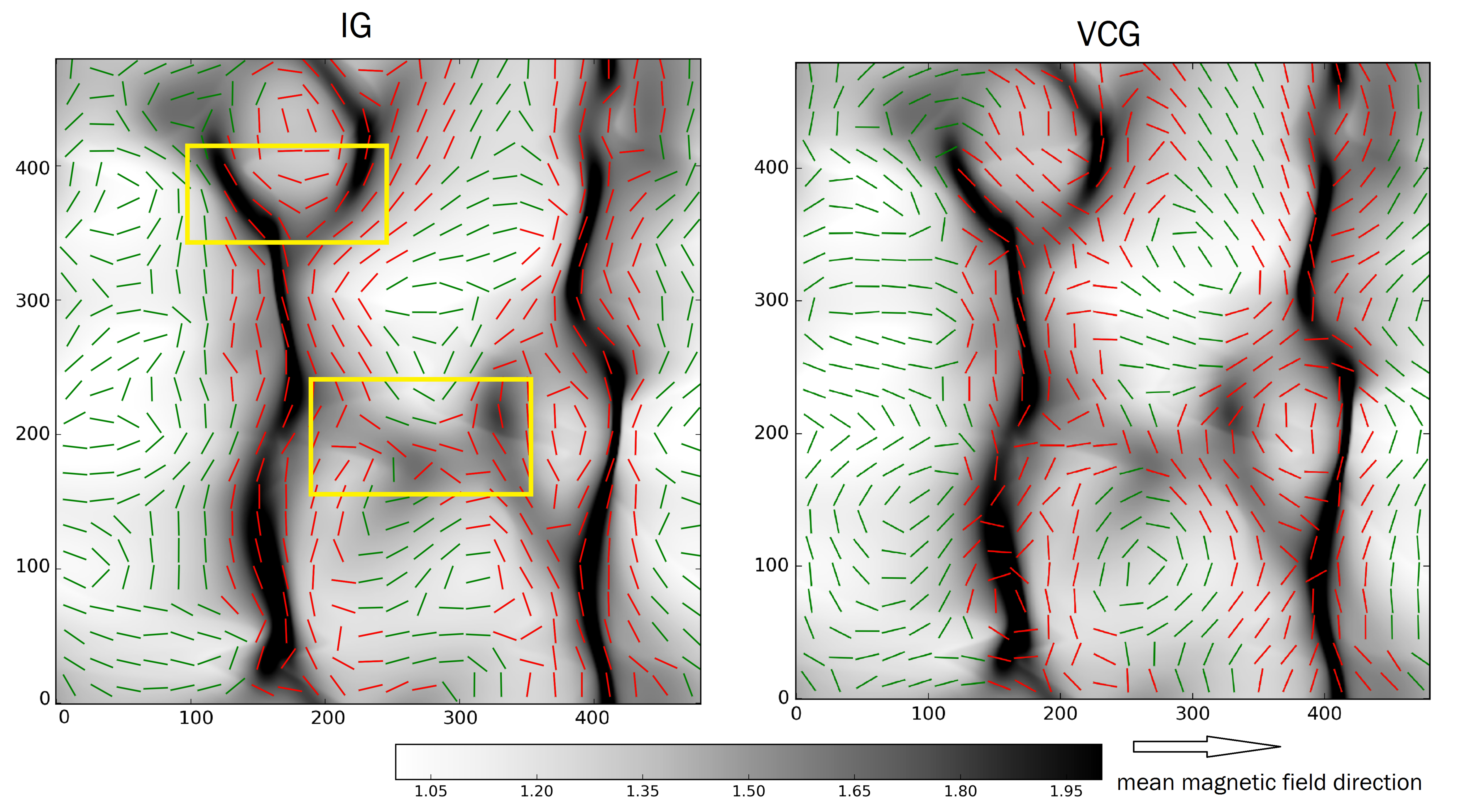}
\includegraphics[width=0.96\textwidth]{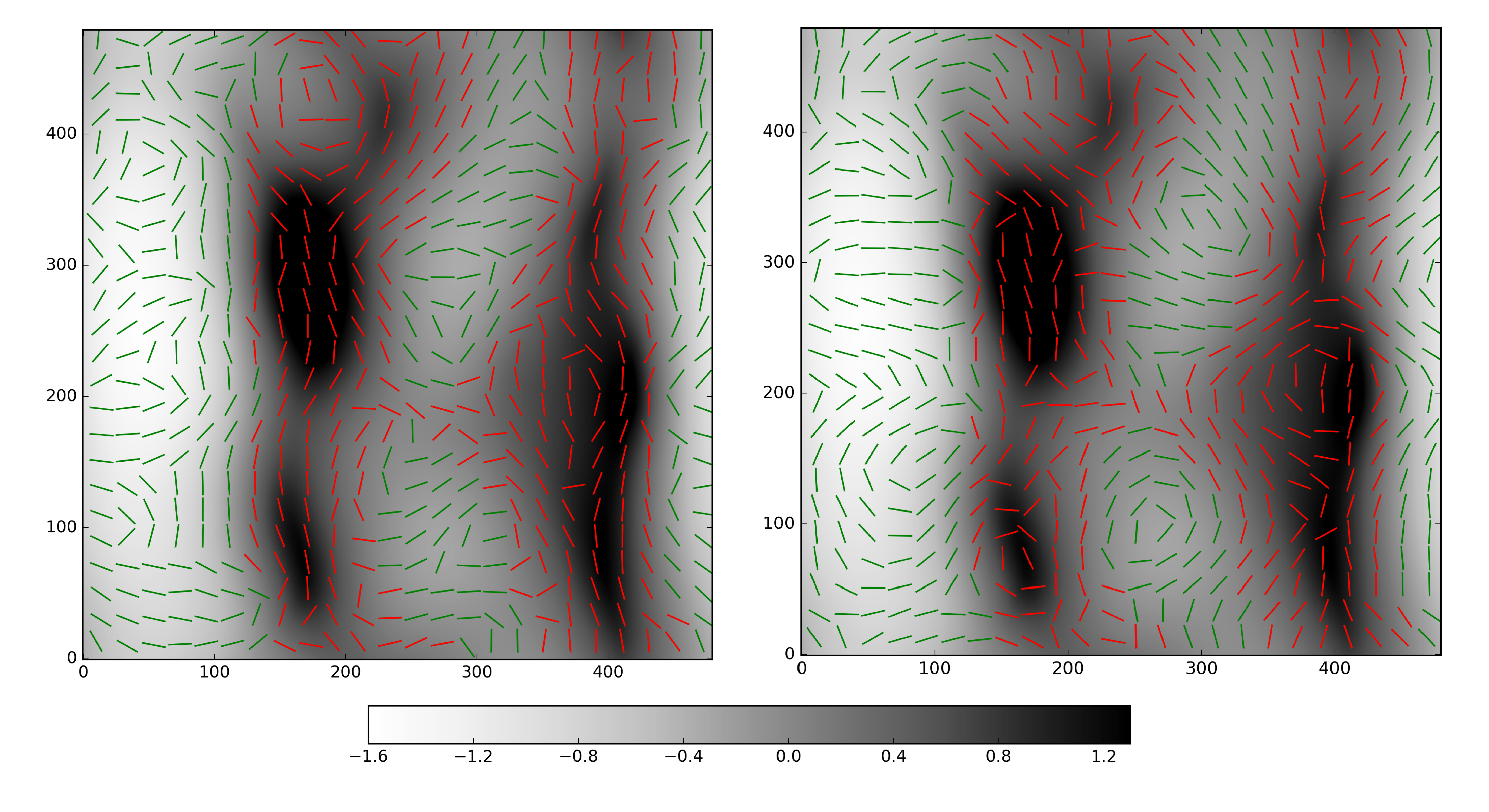}
\caption{\label{fig:grav-gpot} Density (left) and velocity (right) gradients overlaid with the logarithm of densities (upper row) and gravitational energy(lower row). The color of the vector specifies whether the block's gravitational energy is bounded (red vector) or not (green vectors). This system is a very strong field simulation A1, where the magnetic field is {\it always} pointing to the right. The yellow boxes show the region that the density structures are not perpendicular to magnetic field. Block size is 20}
\end{figure*}

It was first reported in \cite{Soler2013} that intensity gradients get parallel to magnetic field when density arrives some threshold $n_T \sim 50 \langle n\rangle $ in the case of strong magnetization media ($\langle B_0 \rangle=10.97$, $\beta=0.1$), and $n_T \sim 500 \langle n \rangle$ in the case of moderately magnetized media ($\langle B_0 \rangle=3.47$, $\beta=1.0$). Moreover, it was claimed that the gravity has only a slight effect on changing the threshold. A similar threshold is also reported when the method is applied for observations \citep{Soler2017a} with the order of neutral hydrogen density at $\langle N_H\rangle \sim 10^{22.2} cm^{-2}$.  We believe that while high column densities are related to self-gravity, whether a particular region is collapsing is determined by the local interactions between magnetic field, turbulence and gravity, and not just density of gas itself. In fact, Figure \ref{fig:grav-dvb} already illustrates the rotation of gradient vectors depends on local physics rather than a particular universal density threshold. To verify our statement, we select some cores from the snapshot on a later-stage self-gravitating simulation and examine the alignment of three-dimensional velocity and density gradients to magnetic field with respect to the density and gravitational energy contours. 

The upper row of Figure \ref{fig:grav-gpot} shows both density and velocity gradients in a slice of the self-gravitating simulation cube overlaid by logarithm of densities.  It is obvious that, while the major filamentary features are perpendicular to magnetic field, there are several density structures that are not perpendicular to magnetic fields. Those are included in a yellow box in Figure \ref{fig:grav-gpot}. Moreover, it can be clearly seen that the red vectors, which represents the region that is self-gravitating, are more likely to be parallel to magnetic field. It is also more clearly seen on the lower row of Figure \ref{fig:grav-gpot}, with the overlaid map be the gravitational energy\footnote{In a periodic simulation, the correct form of gravitational energy should be:
\begin{align}
E_G &= \int d^3 r\frac{(\nabla \Phi(r))^2}{8\pi}
\end{align}
instead of the traditional $E_G = \int d^3r \rho(r) \Phi(r)$. The form of gravitational energy is just the squared gradient amplitude of the gravitational potential $\Phi$.}, the alignment of density and velocity gradient vectors tend to be re-rotated when the energy value arrives $\sim 0.5 pc^2 Myr^{-2}$, but a universal threshold for re-rotation is not found for the whole cube. In fact, Figure \ref{fig:Gravity-AM} already illustrates very clearly how self-gravity should change the alignment of gradient vectors to magnetic field {\it dynamically} when self-gravity is present. We therefore conclude that the 90 degree rotation of the gradient vectors depends more than on the density threshold.

\section{Observational examples}
\label{sec:5}

\subsection{Using GALFA-HI to trace prompting regions for star formation.}
\label{subsec:gadfa}
\begin{figure*}[t]
\centering
\includegraphics[width=0.90\textwidth]{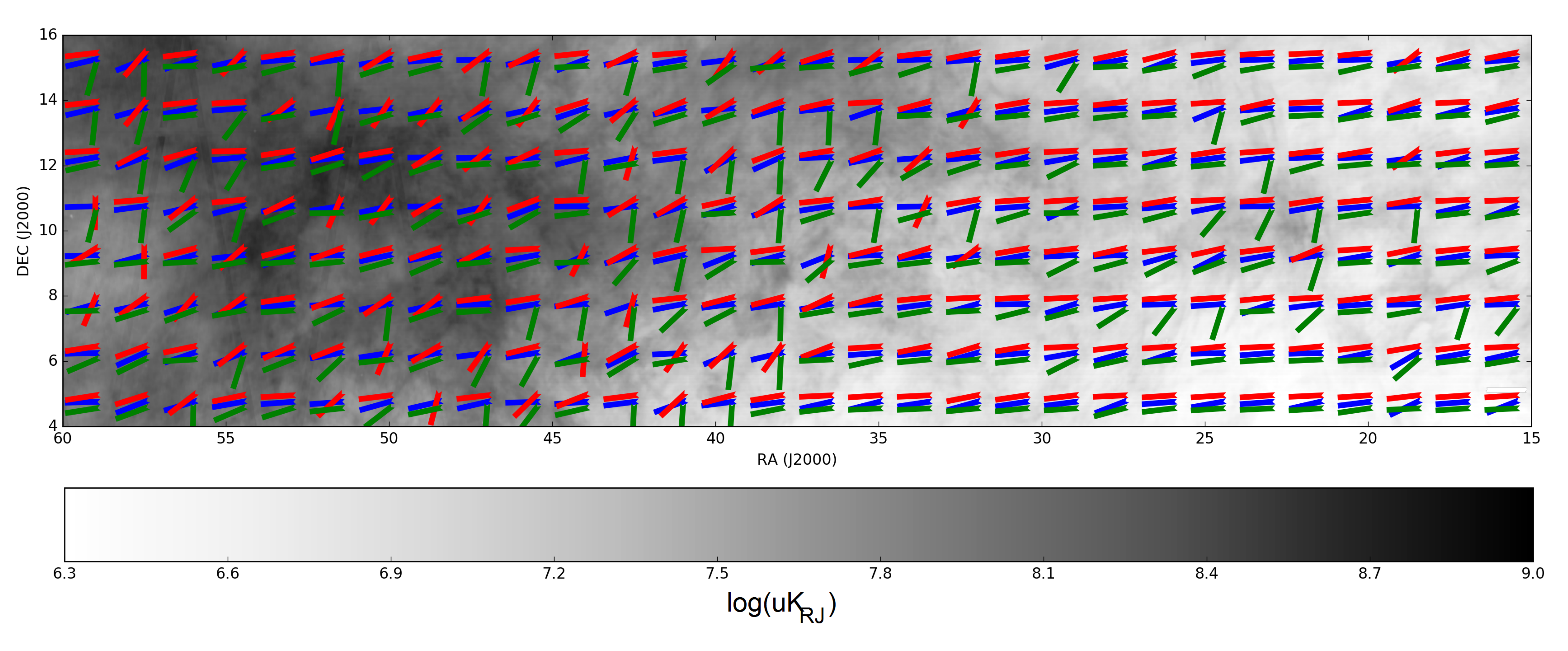} 
\caption{\label{fig:obs-data-vg} Gradients obtained using GALFA-HI survey compared with the Planck polarization data. Here the magnetic field inferred from Planck polarization is shown using blue, while the red and green "vectors" correspond to {\it rotated} the VCGs and the IGs, respectively. }
\end{figure*}

We first illustrate the aforementioned effect by the data from the Galactic Arecibo L-Band Feed Array HI Survey (GALFA-HI). We illustrate the magnetic field tracing with the VCGs and compare it to the {\it Planck} dust thermal polarization data, in which polarization is  perpendicular} to the local magnetic field direction see \citep{2007JQSRT.106..225L,Andersson2015InterstellarAlignment}. Our chosen region from GALFA-HI survey data spans in right ascension from $15^o$ to $60^o$ and in declination from $4^o$ to $16^o$, which is around the position of the South Pole. The velocity channel width is 0.18 km/s, and angular resolution of $3.35'$.  We also plotted the 353GHz dust thermal polarization data obtained by the {\it Planck} satellite's High Frequency Instrument (HFI).\footnote{We use the {\it planckpy} module to extract polarization data in a particular region with J2000 equatorial coordinate: (https://bitbucket.org/ezbc/planckpy/src).}

The VCGs corresponding to the right hand side ($RA\le 35^o$) of Figure \ref{fig:obs-data-vg} have been already compared to the magnetic field directions given by polarization in YL17. Here, we also show the IGs. It is evident that the IGs are worse tracers of magnetic field direction compared to the VCGs. However, the misalignment of the VCGs and the IGs is informative. On the basis of our numerical experiments we can argue that the points where the VCGs and the IGs are not aligned correspond to shocks in the media.

We also see a noticeable change of alignment at the left hand side of Figure \ref{fig:obs-data-vg}. There the density is increased and one may except that self-gravity to become more important. We observe that the VCGs get less aligned with the magnetic field when the column density goes up. The misalignment is even stronger for the IGs. We think that this can, indeed, be interpreted as the effect of self-gravity.  

\subsection{Using full-sky HI data plus synchrotron to trace collapsing regions}

We also would illustrate the power of using the VCGs together with the SIGs. Due to the fact that the full-sky measurement from GALFA-HI is not yet publicly available, we instead pick the HI4PI survey \citep{HI4PI} as illustration, which is based on data from the recently completed first coverage of the Effelsberg-Bonn HI Survey (EBHIS) and from the third revision of the Galactic All-Sky Survey (GASS), to compare with the {\it Planck} foreground Synchrotron Intensity.  With a velocity channel separation of $\delta v=1.29 kms^-1$ and angular resolution of $16'.2$, the HI4PI survey can be used to illustrate the potential of the gradient techniques. To compare also with magnetic field directions from polarization, we acquire the magnetic field direction from the 353Ghz polarization map used in Section \ref{subsec:gadfa}

\begin{figure*}[t]
\centering
\includegraphics[width=0.90\textwidth]{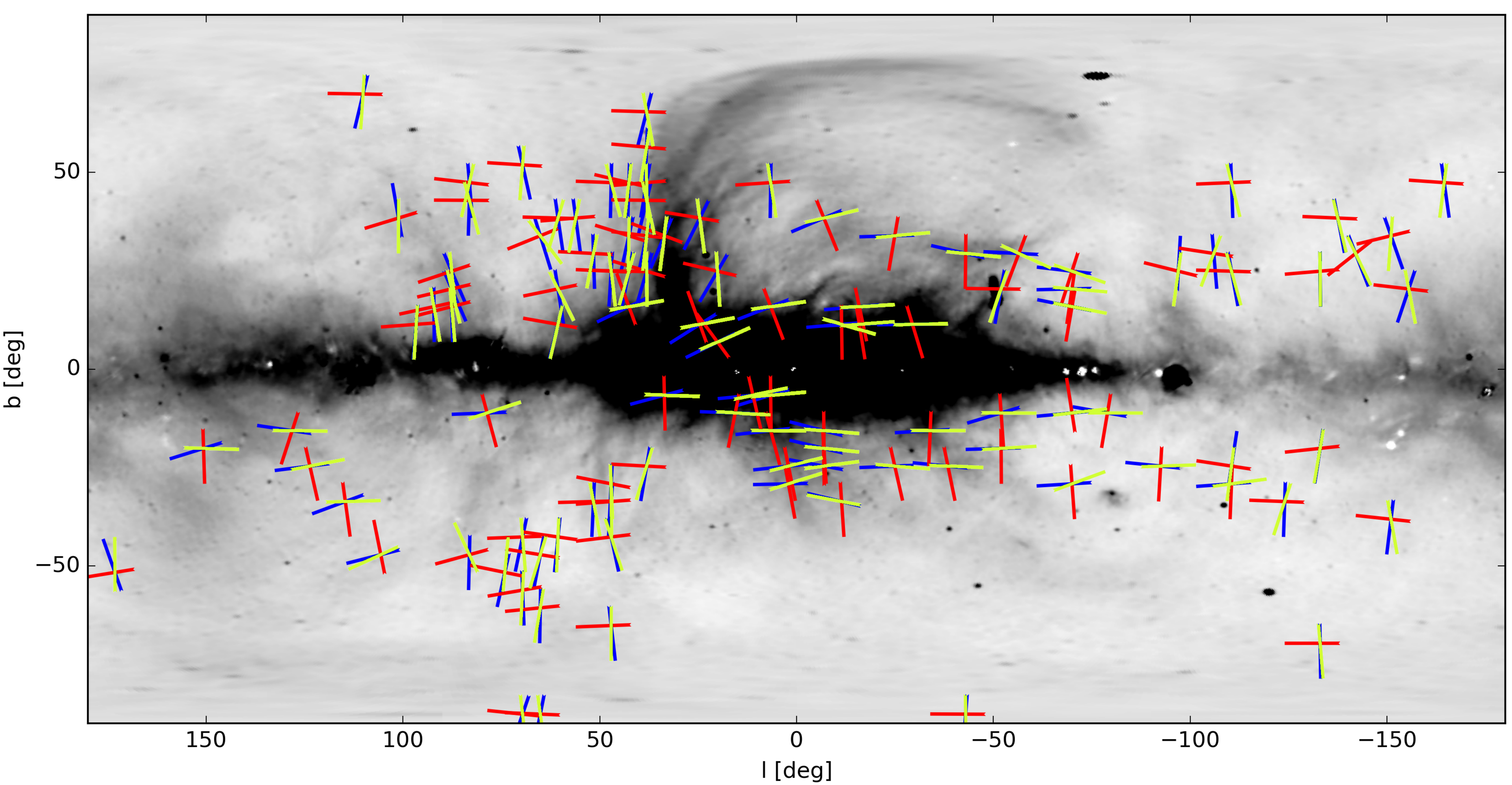} 
\caption{\label{fig:obs-HI-SIG} {\it Rotated} VCGs (red) obtained using HI4PY survey compared with SIGs from Planck synchrotron intensity data (yellow) and magnetic field inferred from Planck 353GHz HFI polarization data (blue), which only the regions with {\it rotated} VCGs being perpendicular to SIGs and magnetic field are shown, which suggests the possibility of the sites for self-gravitation. Only vectors with relative angle larger than $80^o$ have been kept. }
\end{figure*}

We merge the 180 $20^o \times 20^o$ patches available from HI4PI archive and compute the velocity centroid map using information from all channels. From these the VCGs are calculated with the recipe from YL17. On the other hand, the synchrotron foreground intensity map used in \cite{LYLC17} is reused here, with a resolution of $5'$. We calibrate the block size of each of these maps to make sure they are showing the exactly same region with same amount of sub-block averaged vectors for easy comparison. The SIGs, magnetic field from 353Ghz Planck polarization and {\it rotated} the VCGs are then showed in Figure \ref{fig:obs-HI-SIG}. In this figure, we only keep regions that the relative angle between the SIGs and magnetic field and {\it rotated} VCGs is larger than $80^o$. The regions with vectors are suggesting the possibility of the sites for self-gravitation by the two indicators.

The reader should keep in mind that, the region that emits synchrotron radiation can be different from that are responsible for 21 cm emission from  HI. Indeed, the extend of the galactic synchrotron halo is significantly larger than that of the  HI disk. Nevertheless it is encouraging that 
Figure \ref{fig:obs-HI-SIG} 
the directions of the SIGs is well correlated with the magnetic field direction given by the Planck 353 GHz data. In addition, Figure \ref{fig:obs-HI-SIG} demonstrates a good correspondence between the high density region and the region with SIGs-VCGs aligned perpendicularity to each other. Further studies of the synergy of the VCGs and the SIGs is required.

\section{Comparison of VCGs to measuring anisotropy of correlations}
\label{sec:6}
\begin{figure}[h]
\centering
\includegraphics[width=0.48\textwidth]{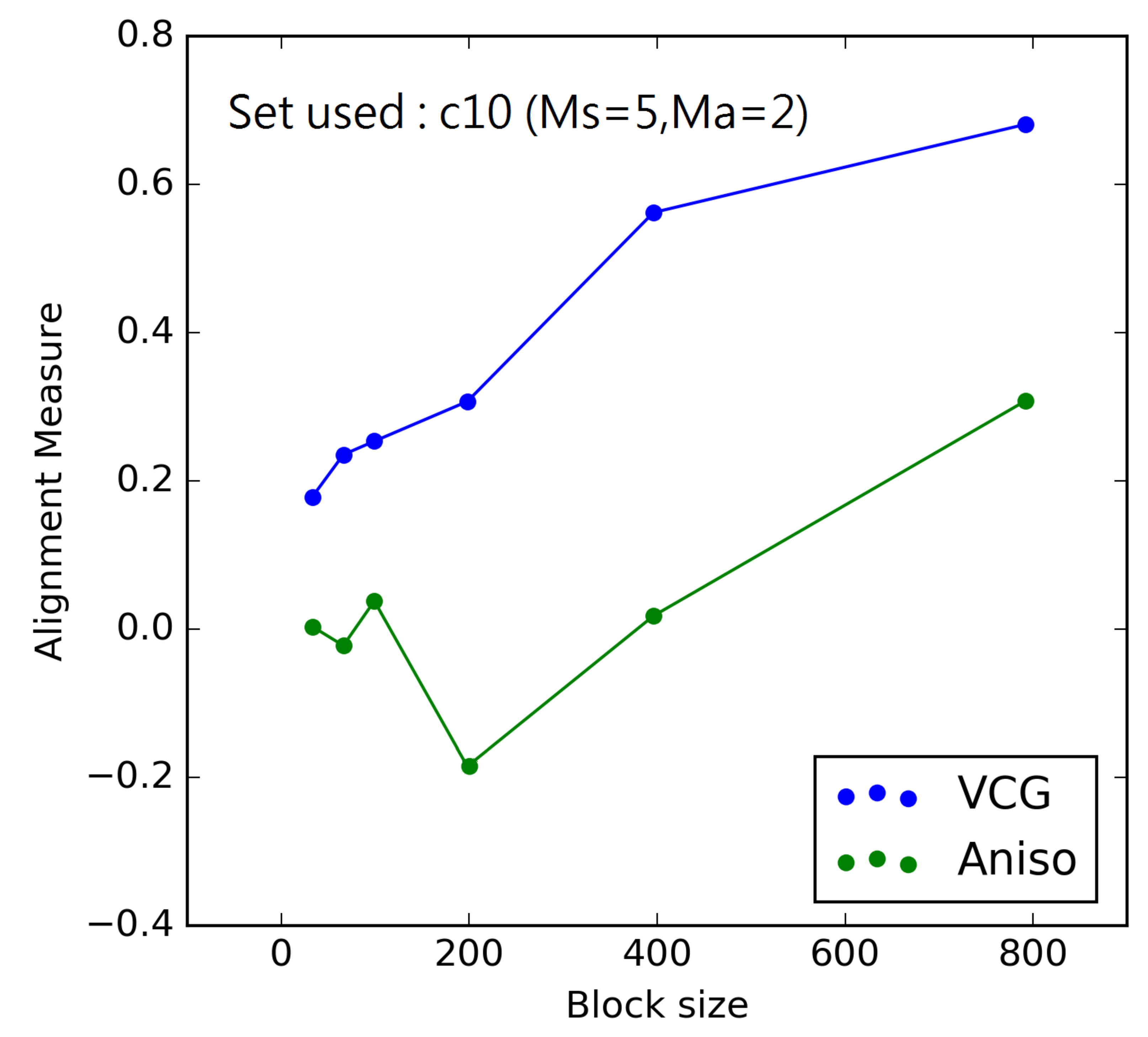}
\caption{\label{fig:8} AM of VCGs compared to that of CFA. The correlation function anisotropy of velocity centroid and intensity is calculated within the block respectively. }
\end{figure}

The anisotropic nature of MHD turbulence induces the anisotropy of observed velocity correlations. This was first demonstrated in \cite{2002ASPC..276..182L} where the velocity correlations within a channel map obtained with synthetic observations based on the result of 3D MHD turbulence were shown. The anisotropy of this map was suggested by \cite{2002ASPC..276..182L} to be used to study the magnetic field direction. Later the anisotropies were studied using centroids in \cite{EL05}. Subsequently, the anisotropies were studied using the Principal Component Analysis (PCA) in \cite{2008ApJ...680..420H}, which is the way of studying anisotropies that equivalent to the aforementioned centroid technique.

The technique of using velocity centroid correlation function anisotropy (CFA) to probe the direction of magnetic fields was further elaborated in \citep{EV11} and was presented as a technique to study magnetization, i.e. the Alfven Mach number $M_A$ of turbulence. Potentially, one can apply the technique to smaller patches of sky to map the local direction of magnetic fields. However, applying of the correlations assumes that the averaging over an ensemble of realizations is available (see \citealt{1975mit..bookR....M}). In practical terms this means that the volume averaging is performed over the patch of the sky much larger than the correlation scale of the eddies. This contradicts the idea of {\it local} measurements and the corresponding weakness of  this approach of detailed tracing magnetic fields is illustrated below. 

The CFA technique relies on the statistics of correlation over a large patch of data to give a clear anisotropy. However, when the patches used for correlation are not large, the shape of the correlation function becomes uncertain, hence a prediction from such calculation fails to trace the local direction of magnetic fields.
In other words, it is difficult to apply the CFA for probing the detailed structure of magnetic fields. In fact, our study of Synchrotron Intensity Gradients (SIGs) in LYLC17 showed the $AM$ of SIGs is significantly better than that of the CFA on smaller scales, and becomes comparable only at the largest scales. 

To perform the same test as in LYLC17, we selected a cube and performed a $AM$-block test for velocity centroids and intensity maps.  Guided by \cite{EV11} we developed an algorithm for calculating the CFA.  As the correlations get elongated along the magnetic field direction, the maximum gradient is perpendicular to it. We calculate the gradients within the correlation function space, and rotate 90 degrees to acquire the field direction prediction. To compare the sub-block average vectors from our recipe, we use the same block to calculate the correlation function.  

Figure \ref{fig:8} shows the comparison of $AM$ for these two methods. While in principle the alignment from the CFA is improving as the block size increases, the VCGs still shows a better alignment in large scales. In small scales, the CFA does not show any preferential alignment. In contrast, the alignment of the VCGs remains positive over all scales. That indicates the superiority of using the VCGs over the CFA in {\it all} scales. At the same time, the CFAs have been studied more theoretically (see \citealt{2017MNRAS.464.3617K} ) with the relation between the degree and anisotropy and composition of MHD turbulence in terms of Alfven, slow and fast modes being analytically established. Thus we believe in the synergy of using both the VCGs and the CFAs for studying turbulence and magnetic field in the ISM.

\section{Discussion}
\label{sec:7}
\subsection{Comparison with the earlier work}

The use of 3D density and intensity gradients to gauge the relative importance of magnetic field, i.e the relative magnetization on a particular region, was suggested in the influential study by \cite{Soler2013} on the basis of numerical studies.  The Histogram of Relative Orientations (HRO) based on this study has been recently applied to  the Vela C molecular cloud data. \cite{Soler2017a}. HRO is the measure of the distribution of relative angle between the intensity gradients and the projected magnetic field inferred using dust polarization provides an insight whether whether the region has strong magnetic field or not. Our approach to the gradient techniques, including the IGs is different, however. Our studies based on the modern understanding of MHD turbulence theory that established that eddies will be anisotropic and well aligned with the local direction of magnetic field in case of sub-Alfvenic turbulence and the small scales in super-Alfvenic turbulence. The smallest eddies that are best aligned with the local direction of magnetic field are responsible for the largest velocity gradients, which provides a way of measuring the local magnetic field. Therefore our studies 
are aimed at obtaining {\it point-wise} measures of magnetic field as well as other measures that can affect this alignment, e.g. shocks and regions of gravitational collapse (see GL17, YL17, LYLC17, LY17). In terms of magnetic field tracing we rely mostly on the VCGs and SIGs, while stress the synergetic use of these techniques when combined with the polarimetry and the IGs. The former shows significantly better alignment with magnetic field compared to the latter. Nevertheless, we stress that in some cases when the intensities, e.g. total line emission intensity or dust emission intensity, are only available, the IGs can also be used as point-wise magnetic field tracers at least in the diffuse low sonic Mach number media. In particular, in this paper we introduced the ways of separating the regions affected by shocks and those where the density is expected to obey the GS95 relations.\footnote{It was shown in Beresnyak, Lazarian \& Cho 2005 that filtering the regions where the density is enhanced by shocks one can observe the GS95 scaling for less disturbed regions.} 

On the utilization of the gradient method, our approach is to {\bf predict} magnetic field direction through the quantitative method of {\it sub-block averaging} (YL17), which we also apply to calculating the IGs. While the method does not guarantee the resultant vectors are always perfectly align to the local magnetic field direction, the sub-block Gaussian divergence can already tell the accuracy of our technique, and our previous work e.g. in YL17 as well as our present work suggests that an appreciable alignment is available with both the VCGs and the IGs. On the other hand, without the procedure we proposed in YL17, the approach in \cite{Soler2013} relies on the presence of polarization data to judge the importance of magnetic field in s region, which, technically, is treating the whole map as a single big block. We also note that the HROs does not provide any kinds of tracing of magnetic field.  On the other hand, given a good alignment between the magnetic fields and the gradients, they can in principle be used as proxies of polarization within the Chandrasekhar-Fermi (CF) method (GL17, YL17). We have demonstrated that the simultaneous usage of the IGs and the VCGs could also identify the regions where other physical effects that shape the ISM take place, e.g. shocks and gravitational collapse. In fact, in the present study we show the relative alignment between IGs and VCGs can actually shed the stage of collapse. 

Another important study that preceded out work on gradients was the study of intensity filament alignment with magnetic field that was observed in HI channel maps in \cite{Clark2015NeutralForegrounds}. As we mentioned earlier, on the basis of the theoretical work in \cite{LP00} (see also \citealt{L01,CL09,2017MNRAS.464.3617K}) we conclude that the filaments observed in the thin channel maps can arise from velocity crowding in the velocity space, i.e. be the velocity caustics, rather than the actual density filaments. In LY17 we compare the tracing of magnetic field that can be obtained with the velocity channel filaments with that produced by gradients calculated with the intensities within the velocity channel maps.

\subsection{Advantages of the VCGs and the synergy of VCGs \& IG }

Even though we show that VCGs are superior in terms of magnetic field tracing than IGs, the IGs are relatively more sensitive to shocks, which not only change the orientation of the IGs in respect to magnetic field, but also increase the amplitude of the IGs. As a result, removal of pixels corresponding to the IGs with large amplitude can be mitigate the effect of the shocks for the VCGs.  

We view the comparison of the relative alignment between the IGs and the the VCGs as a viable tool to observational studies of of shock and regions of gravitational collapse.  We expect a decrease of ability of the IGs on having coherent alignment to VCGs. Figure \ref{fig:VCGIG-Ms} shows our numerical test on the relative alignment between the IGs and the VCGs as $M_s$ increases.  A decrease of the AM is clearly seen in the figure. A similar tendency of the change of the relative alignment is observe in self-gravitating regions as illustrated by Figure \ref{fig:Gravity-AM}, where the IGs are relatively well aligned with the VCGs in very diffuse region, but then becoming less aligned in intermediate stage of self-collapse to become again aligned at the last stage of the gravitational collapse. We provide a summary of these features in Table \ref{tab:3}. Although these two effects may interfere to each other, the synergy of using both the VCGs and the IGs a very important insight into the interstellar dynamics even in the absence of polarimetry data.  

\begin{table}
 \centering
 \label{tab:3}
 \begin{tabular}{c c c c}
  Stage of collapse & Diffuse  & Intermediate & Late-time \\ \hline \hline
  VCGs vs IGs & $\parallel $ & $\bot$ & $\parallel$  \\ \hline
 \end{tabular}
  \caption {A summary on the expected directions of the relative alignment between VCGs and IGs}
\end{table}

In a subsequent study on observational studies on the same Vela C region (Yuen et al. in prep), we not only be able to trace the direction of magnetic field in appropriate scales with very high $AM\sim 0.7$, but also locating the possible self-gravitating region using the numerical result in this work. That suggests the applicability of our technique for magnetic field tracing even within extreme environments. 
\subsection{How should one deal with highly super-Alfvenic cases?}
\label{subsec:superalf}

 In this work, we investigate a number of different regimes for the sonic Mach number $M_s$, but the explored range of the Alfven Mach numbers $M_A$ is limited to sub- and trans-Alfvenic cases only. While our series of studies (YL17, LYLC17, this work, \citealt{LY17}) shows the power of our gradient technique in different circumstances, one very important regime that is yet to explore is the highly super-Alfvenic cases.

On the other hand, the expectation from turbulence theory (GS95, LV99) also make the tracing through gradients more complicated than sub- and trans-Alfvenic cases. Only sufficient small scales $l_A = L_{inj} M_A^{-3}$ will the turbulent eddies become anisotropic. In fact, the wavenumber removal suggested in Section \ref{subsec:3.2} should be able to reveal the alignment of the VCGs to underlying magnetic field provided that $l_A$ is known. The practical procedures for the actual data handling will be discussed elsewhere.

\section{Conclusion}
\label{sec:8}
Our paper explores the use of the velocity centroid gradients (VCGs) for tracing magnetic fields and its synergy with other gradient techniques and polarimetry for probing the ISM processes. We tested how VCGs, intensity gradients (IGs) and synchrotron intensity gradients (SIGs) behave under different physical conditions, including a change of sonic and Alfvenic Mach numbers and in the presence of self-gravity.  We show that in the diffuse media the VCGs trace the magnetic field well and showed that in combination with polarimetry they can also be efficient in identifying the regions dominated by self-gravity. Our results in the paper can be briefly summarized:

\begin{enumerate}
\item In the diffuse media the VCGs trace the magnetic field well and in combination with polarimetry or SIGs they can also be efficient in identifying the regions dominated by self-gravity.
\item  The deviations of the direction of the VCGs and the IGs can signify the existence of shocks and can be used as a ways to obtain the turbulence sonic Mach number.
\item New ways to improve the performance of the VCGs and IGs tracing magnetic fields were identified. They include 
\begin{enumerate}
\item Removing pixels with higher IG amplitudes. 
\item Removing the contribution of lower wave number. 
\end{enumerate}
\item For self-gravitating regions the alignment of the VCGs changes:
\begin{enumerate}
\item In the case of dynamically important magnetic field, the unconstrained falling of matter along magnetic fields lines makes the gradients aligned with magnetic fields. Therefore the change of the relative orientation of polarization directions and those determined by the VCGs can provide the observational signature of the self-gravitating collapse within molecular clouds. 
\item The time scale for the response of the IGs  to the gravity is faster than the VCGs in a collapsing cloud, which shed the possibility of determining the stage of collapse by tracing the relative alignment of the IGs to the VCGs in the vicinity of high density regions. This opens a way to identify the regions of gravitational collapse by observing the increase of misalignment of the IGs and the VCGs for intermediate densities with the return to their alignment for higher densities. 
\item Combining VCGs and SIGs explored earlier in \cite{LYLC17}, it is possible to locate self-gravitating regions by examining the relative orientation of the two gradients. 
\item The gravity, rather than the density threshold determine the change of the direction of gradient vectors in self-gravitating regions.
\end{enumerate}
\item The VCGs can provide more detailed magnetic field tracing compared to the correlation function anisotropy technique as suggested in Lazarian et al. (2002). 
\item We illustrate our technique by analyzing the observation data:
\begin{enumerate}
\item For the nearby region of GALFA-HI used in \cite{YL17} to the combination of the VCGs and the IGs reveals shocks.
\item The full-sky HI4PI HI data together with Planck synchrotron foreground intensity data and Planck dust polarization data locate the possible self-gravitating regions.
\end{enumerate} 

\end{enumerate}

\acknowledgements The stay of KHY at UW-Madison is supported by the Fulbright-Lee Hysan research fellowship and Department of Physics, CUHK. AL acknowledges the support of the NSF grant AST 1212096, NASA grant NNX14AJ53G as well as a distinguished visitor PVE/CAPES appointment at the Physics Graduate Program of the Federal University of Rio Grande do Norte, the INCT INEspao and Physics Graduate Program/UFRN.












\end{document}